\pdfoutput=1
\RequirePackage{ifpdf}
\ifpdf % We are running pdfTeX in pdf mode
\documentclass[pdftex]{sigma}
\else
\documentclass{sigma}
\fi

\def\asl{\operatorname{sl}}

\newcommand{\ri}{\mathrm{i}}
\newcommand{\rd}{\mathrm{d}}\newcommand{\rj}{\mathrm{j}}
\def\ad{\operatorname{ad}}
\def\tr{\operatorname{tr}}
\def\const{\operatorname{const}}
\def\Ad{\operatorname{Ad}}
\def\diag{\operatorname{diag}}
\def\id{\operatorname{id}}
\def\SU{\operatorname{SU}}
\def\so{\operatorname{so}}
\def\u{\operatorname{u}}
\def\su{\operatorname{su}}
\def\im{\operatorname{im}}
\def\pd#1,#2{\frac{\partial#1}{\partial#2}}

\begin{document}

\allowdisplaybreaks

\renewcommand{\PaperNumber}{087}

\FirstPageHeading

\ShortArticleName{Geometry of ${\rm sl}(n,{\mathbb C})$ Recursion Operators}

\ArticleName{Geometric Theory of the Recursion Operators for the Generalized Zakharov--Shabat System in Pole Gauge on the Algebra
$\boldsymbol{{\rm sl}(n,{\mathbb C})}$
with and without Reductions}

\Author{Alexandar B. YANOVSKI~$^\dag$ and Gaetano VILASI~$^\ddag$}

\AuthorNameForHeading{A.B.~Yanovski and G.~Vilasi}

\Address{$^\dag$~Department of Mathematics \& Applied Mathematics, University of Cape Town,\\
\hphantom{$^\dag$}~Rondebosch 7700, Cape Town, South Africa}
\EmailD{\href{mailto:Alexandar.Ianovsky@uct.ac.za}{Alexandar.Ianovsky@uct.ac.za}}

\Address{$^\ddag$~Dipartimento di Fisica, Universit\`a degli Studi di Salerno, INFN, Sezione di Napoli-GC Salerno,\\
\hphantom{$^\ddag$}~Via Ponte Don Melillo, 84084, Fisciano (Salerno), Italy}
\EmailD{\href{mailto:vilasi@sa.infn.it}{vilasi@sa.infn.it}}

\ArticleDates{Received May 17, 2012, in f\/inal form November 05, 2012; Published online November 16, 2012}

\Abstract{We consider the recursion operator approach to the soliton equations related to the generalized Zakharov--Shabat system on the algebra $\asl(n,\mathbb{C})$ in pole gauge both in the general position and in the presence of reductions. We present the recursion operators and discuss their geometric meaning as conjugate to Nijenhuis tensors for a Poisson--Nijenhuis structure def\/ined on the manifold of potentials.}


\Classification{35Q51; 37K05; 37K10}

\vspace{-2mm}

\section{Introduction}

The theory of nonlinear evolution equations (NLEEs) of soliton type (soliton equations or completely integrable equations) has developed considerably in recent decades. Interest in it is still  big and there is a wide variety of approaches to these equations.  However, some properties are fundamental to all approaches and one is that these equations admit the so called Lax representation, namely $[L,A]=0$.  In the last expression $L$ and $A$ are linear operators on~$\partial_x$,~$\partial_t$ depending also on some functions $q_{\alpha}(x,t)$, $1\leq \alpha\leq s$ (called `potentials') and a spectral para\-me\-ter~$\lambda$. Since the Lax equation $[L,A]=0$ must be satisf\/ied identically in $\lambda$, it is equivalent to a~system (in the case when $A$ depends linearly on $\partial_t$) of the type
\begin{gather}\label{eq:evp}
(q_{\alpha})_t=F_{\alpha}(q,q_x,\dots), \qquad \mbox{where} \quad   q=(q_{\alpha})_{1\leq \alpha\leq s}.
\end{gather}

In most of the approaches the linear problem $L\psi=0$ (auxiliary linear problem) remains f\/ixed and the evolution equations (of a certain form) that can be obtained by changing the opera\-tor~$A$ are considered. The hierarchies of equations we obtain by f\/ixing $L$ are called the nonlinear evolution equations (NLEEs), or soliton equations, associated with (or related to) $L$ (or with the linear system $L\psi=0$). The hierarchies usually are named for some of the remarkable equations contained in them. The schemes according to which one can calculate the solutions to the soliton equations may be very dif\/ferent, but the essential fact is that the Lax representation permits one to pass from the original evolution def\/ined by the system of equations~\eqref{eq:evp} to the evolution of some spectral data related to the problem $L\psi=0$. Since f\/inding the spectral data evolution usually is not a problem, the principal dif\/f\/iculty is to recover the potentials from the spectral data. This process is  called  the inverse scattering method, which is described in detail in the monographs~\cite{FadTakh87,GerViYa2008}.

The generalized Zakharov--Shabat system (GZS system) we see below is one of the best known auxiliary linear problems. It can be written as follows
\begin{gather}\label{eq:GZS}
L\psi=\left( \ri\partial_x + q(x) -\lambda J \right) \psi = 0.
\end{gather}
Here $q(x)$ and $J$ belong to a f\/ixed simple Lie algebra $\mathfrak{g}$ in some f\/inite-dimensional irreducible representation. The element $J$ is regular, that is the kernel of $\ad_J$ ($ \ad_J(X )\equiv [J,X]$, $X \in \mathfrak{g}$) is a Cartan subalgebra
$\mathfrak{h} \subset \mathfrak{g}$. The potential $q(x)$ belongs to the orthogonal complement ${\mathfrak{h}}^{\perp}$ of $\mathfrak{h}$ with respect to the Killing form
\begin{gather}\label{1.6}
\langle X, Y \rangle = \tr(\ad_X \ad_Y), \qquad X, Y \in \mathfrak{g},
\end{gather}
and therefore $q(x)=\sum\limits_{\alpha\in \Delta} q_{\alpha}E_{\alpha}$ where $E_{\alpha}$ are the root vectors; $\Delta$ is the root system of $\mathfrak{g}$. The scalar functions $q_{\alpha}(x)$ (the `potentials') are def\/ined on $\mathbb{R}$, are complex valued, smooth and tend to zero as $x \to \pm\infty$. We can assume that they are Schwartz-type functions. The classical Zakharov--Shabat system is obtained for $\mathfrak{g} = \asl(2,\mathbb{C})$, $J=\diag(1,-1)$.
\begin{remark}
We assume that the basic properties of the semisimple Lie algebras (real and complex) are known and we do not give def\/initions of all the concepts related to them. All our def\/initions and normalizations coincide with those in \cite{GoGr}.
\end{remark}
\begin{remark}
When generalized Zakharov--Shabat systems on dif\/ferent algebras are involved we say that we have a generalized Zakharov--Shabat $\mathfrak{g}$-system (or generalized Zakharov--Shabat on~$\mathfrak{g}$) to underline the fact that it is on the algebra $\mathfrak{g}$. When we work on a f\/ixed algebra its symbol is usually omitted.
\end{remark}
Here we may mention also that, in case when the element $J$ is complex, the problem \eqref{eq:GZS} is referred as a Caudrey--Beals--Coifman system \cite{BeCo84} and only in  the case when~$J$ is real it is called a generalized Zakharov--Shabat system.  The reason for the name change is that the spectral theory of $L$ is of primary importance for the development of the inverse scattering techniques for~$L$ and  the cases when~$J$ is real or complex are quite dif\/ferent from the spectral viewpoint, see for example~\cite{G86,GYa94} in which the completeness of the so-called adjoint solutions of~$L$ when~$L$ is considered in an arbitrary faithful representation of the algebra $\mathfrak{g}$ is proved. Referring for the details to the above work we simply remind the reader that the adjoint solutions of $L$ are functions of the type $w=mXm^{-1}$ where $X$ is a constant element from $\mathfrak{g}$ and $m$ is a fundamental solution of $Lm=0$. Let us denote the orthogonal projector (with respect to the Killing form~\eqref{1.6}) on ${\mathfrak{h}}^{\perp}$ by $\pi_0$. Then, of course, the orthogonal projector on $\mathfrak{h}$ will be equal to $\id-\pi_0$. Further, let us put  $w^{\rm a}=\pi_0w$ and $w^{\rm d}=(\id-\pi_0)w$. One of the most important facts from the theory of GZS systems is that if a suitable set of adjoint solutions $(w_i(x,\lambda))$ is taken, then, roughly speaking, for~$\lambda$ on the spectrum of $L$ the functions $(w_i^{\rm{a}}(x,\lambda))$ form a complete set in the space of potentials.  If one expands the potential over the subset of the adjoint solutions the coef\/f\/icients are one of the possible minimal scattering data sets for $L$. Thus, passing from the potentials to the scattering data can be considered as a sort of Fourier transform, called a generalized Fourier transform. For it $w_i^{\rm{a}}(x,\lambda)$ play the role the exponents play in the usual Fourier transform. They are called generalized exponents or by abuse of language also adjoint solutions. Those familiar with the theory in $\asl(2;\mathbb{C})$ know that originally they were called `squares' of the solutions of $L\psi=0$. This interpretation of the inverse scattering transform was given for the f\/irst time in~\cite{AKNS} and after that has been developed in a number of works, see,  for example, the monographs~\cite{GerViYa2008, IKhKi94} for complete study of $\asl(2,\mathbb{C})$-case and for  comprehensive bibliographies, and~\cite{BeCo84,GYa94} for more general situations.

However, since in this article we shall not deal with the spectral properties of~$L$, we shall call it a generalized Zakharov--Shabat system in all cases.

The recursion operators (generating operators, $\Lambda$-operators) are the operators for which the functions $w_i^{\rm{a}}(x,\lambda)$ are eigenfunctions and therefore for the generalized Fourier transform they play the same role as the dif\/ferentiation operator in the usual Fourier transform method.  For that reason recursion operators play central role in the theory of soliton equations~-- it is a~theoretical tool which, apart from explicit solutions, can give most of the information about the NLEEs \cite{GerViYa2008,ZaKo85}. The theory of these operators is an interesting and developing area. Through them one may obtain:
\begin{itemize}\itemsep=0pt
\item[i)] the hierarchies of  the nonlinear evolution equations solvable through~$L$;
\item[ii)] the conservation laws for these NLEEs;
\item[iii)] the hierarchies of Hamiltonian structures for these NLEEs.
\end{itemize}
It is not hard to f\/ind that the recursion operators related to $L$ have the form, see \cite{GYa94} or the book \cite{GerViYa2008} and the numerous references therein,
\begin{gather}\label{eq:LOR}
\Lambda_{\pm}(X(x))=
\ad_{J}^{-1}\left(\ri \partial_x X+
\pi_0[q,X]+\ri\ad_{q}\int_{\pm
\infty}^{x}(\id-\pi_0)[q(y),X(y)]\rd y \right).
\end{gather}
Here, of course, $\ad_q(X)=[q,X]$ and $X$ is a smooth, rapidly decaying function with values in~$\mathfrak{h}^{\perp}$.

The name `recursion operators' has the following origin. Suppose we are looking for the NLEEs that have Lax representation $[L,A]=0$ with $L$ given in~\eqref{eq:GZS} and $A$ of the form
\begin{gather*}
A = \ri\partial_t + \sum_{k=0}^n \lambda^{k} A_{k}, \qquad
A_n \in \mathfrak{h}, \qquad A_n = \const, \qquad A_{n-1}\in \mathfrak{h}^{\perp}.
\end{gather*}
Then from the condition $[L,A]=0$ we f\/irst obtain $A_{n-1}=\ad_J^{-1}[q,A_n]$ and for $0<k<n-1$ the recursion relations
\begin{gather*}%\label{eq:ecb}
\pi_{0}A_{k-1}=\Lambda_{\pm}(\pi_{0}A_{k}),\qquad
(\id-\pi_{0})A_k=\ri(\id-\pi_{0})\int_{\pm \infty}^x[q,\pi_0A_k](y)\rd y,
\end{gather*}
where $\Lambda_{\pm}$ are as in \eqref{eq:LOR}. This leads to the fact that the NLEEs related to $L$ can be written in one of the following equivalent forms:
\begin{gather}
{\rm a)}\quad  \ri  \ad_J^{-1}q_t +
\Lambda^n_{+} \left(
\ad_J^{-1}   [A_n , q] \right) = 0 \label{eq:EEZSS},\\
{\rm b)}\quad  \ri \ad_J^{-1} q_t +
\Lambda^n_{-} \left( \ad_J^{-1}   [A_n , q] \right) = 0.\nonumber
\end{gather}
\begin{remark}
Strictly speaking, this is not the most general form of the equations solvable through~$L$.  Considering the right-hand side of the equations of the type $\ad_J^{-1}q_t=F_n(q)$ as vector f\/ields, in order to obtain the general form of the NLEEs associated with~$L$ one must take an arbitrary f\/inite linear combination $F$ of the vector f\/ields $F_n$ with constant coef\/f\/icients and write $\ad_J^{-1}q_t=F(q)$.  We refer to~\eqref{eq:EEZSS} as the general form of the equations solvable through $L$ for the sake of brevity.
\end{remark}
There is another important trend in the theory of the  GZS system and consequently for the recursion operators related to it.  It turns out that this system is closely related to another one, called the GZS system in pole gauge (then the system~$L$ we introduced is called GZS system in canonical gauge). In order to introduce it, denote the  group that corresponds to the algebra~$\mathfrak{g}$ by~$G$. Then the system we are talking about  is the following
\begin{gather}\label{eq:GZSpole}
\tilde{L}\tilde{\psi}=\ri\partial_x\tilde{\psi}-S\tilde{\psi}=0, \qquad S\in \mathcal{O}_J
\end{gather}
(with appropriate conditions on $S(x)$ when $x\mapsto \pm\infty$), where  $\mathcal{O}_J$ is the orbit of the adjoint action of the group $G$,  passing through the element $J\in \mathfrak{g}$. A gauge transformation of the type $\psi\mapsto \psi_0^{-1}\psi=\tilde{\psi}$ where $\psi_0$ is a fundamental solution to the GZS system corresponding to $\lambda=0$ takes the system $L$ into the system $\tilde{L}$ if we denote $S=\psi_0^{-1}J\psi_o$. One can choose dif\/ferent fundamental solutions $\psi_0$ and one will obtain dif\/ferent limiting values for $S$ when $x\mapsto\pm\infty$ but usually for $\psi_0$ is taken the Jost soliton that satisf\/ies $\lim\limits_{x\to -\infty}\psi_0=\mathbf{1}$. The GZS system in pole gauge is used as the auxiliary linear problem to solve the equations that are classical analogues of equations describing waves in magnetic chains. For example, in the case of $\mathfrak{g}=\asl(2,\mathbb{C})$ (spin~${1}\slash{2}$), one of the NLEEs related to $\tilde{L}$ is the Heisenberg ferromagnet equation
\begin{gather}\label{eq:HF}
S_t=\frac{1}{2\ri}[S,S_{xx}],\qquad S\in \ri\su(2).
\end{gather}
In the case of $\asl(3,\mathbb{C})$ the linear problem $\tilde{L}$ is related to a classical analog of equations describing the dynamics of spin $1$ particle chains~\cite{BoPo90,Yan93}.

The theory of NLEEs related to the GZS auxiliary system in canonical gauge~($L$) is in direct connection with the theory of the NLEEs related with the GZS auxiliary system in pole gauge~($\tilde{L}$). The NLEEs for both systems are in one-to-one correspondence and are called gauge-equivalent equations. This beautiful construction was used for the f\/irst time in the famous work of Zakharov and Takhtadjan~\cite{ZaTakh79}, in which the gauge-equivalence of two famous equations~-- the Heisenberg ferromagnet equation \eqref{eq:HF} and the nonlinear Schr\"odinger equation was proved.

In fact the constructions for the system $L$ and its gauge equivalent $\tilde{L}$ are in complete analogy. Instead of the f\/ixed Cartan subalgebra $\mathfrak{h}=\ker\ad_J$, we have a `moving' Cartan subalgebra  $\mathfrak{h}_S=\ker\ad_{S(x)}$; a `moving' orthogonal (with respect to the Killing form) complementary space~$\mathfrak{h}_S^{\perp}$ to~$\mathfrak{h}_S$ etc. We have the corresponding adjoint solutions $\tilde{m}=\tilde{\psi} X \tilde{\psi}^{-1}$, where $\tilde{\psi}$ is a fundamental solution of $\tilde{L}\tilde{\psi}=0$ and $X$ is a constant element in~$\mathfrak{g}$. If we denote by $\tilde{m}^{\rm{a}}$ and~$\tilde{m}^{\rm{d}}$ the projections of $\tilde{m}$ on  $\mathfrak{h}_S^{\perp}$ and $\mathfrak{h}_S$ respectively, then  the corresponding recursion operators are constructed using the fact that the functions $\tilde{m}^{\rm{a}}$ must be eigenfunctions for them. The evolution equations associated with the system~\eqref{eq:GZSpole} and gauge-equivalent to the equations~\eqref{eq:EEZSS}  have the form
\begin{gather}\label{eq:EEZSSspole}
-\ri\ad_{S}^{-1}S_t+
(\tilde{\Lambda}_{\pm})^n\tilde{\Lambda}_{\pm}\pi_{S}A_n=0,
\end{gather}
where $\tilde{\Lambda}_{\pm}$ are the recursion operators for $\tilde{L}$ and $\pi_S$ is the orthogonal projector on $\mathfrak{h}_S^{\perp}$. One can see that
\begin{gather*}
\tilde{\Lambda}_{\pm}=\Ad\big(\psi_0^{-1}\big)\circ \Lambda_{\pm}\circ \Ad(\psi_0), \qquad
{\pi}_S=\Ad\big(\psi_0^{-1}\big)\circ \pi_0 \circ \Ad(\psi_0),
\end{gather*}
where $\Ad$ is denotes the adjoint action of the simply connected Lie group $G$ having $\mathfrak{g}$ as algebra.
\begin{remark}
In order to understand why in the hierarchy \eqref{eq:EEZSSspole} appears $\pi_{S}A_n$ one must mention that one can prove that for any constant $H\in\mathfrak{h}$
\[
\Ad\big(\psi_0^{-1}\big)\big(\ad_J^{-1}[q,H]\big)={\pi}_S H.
\]
\end{remark}
So for the GZS system in pole gauge everything could be reformulated  and the only dif\/f\/iculty is to calculate all the quantities that are expressed through $q$ and its derivatives through $S$ and its derivatives.  Though in each particular case the details may be dif\/ferent there is a clear procedure to achieving that goal. The procedure was developed in the PhD thesis \cite{Yan87}, outlined in~\cite{GerYan84,GerYan85} for the $\asl(2,\mathbb{C})$ case, and for more general cases in~\cite{GerYan85b}. In the case $\asl(3,\mathbb{C})$ the procedure has been carried out in detail in~\cite{Yan93}. The theory of the recursion operator for the GZS system in canonical gauge in the presence of the Mikhailov type reductions has been also considered, see for example~\cite{GerGrahMikhVal*2011} for a treatment from the spectral theory viewpoint, or~\cite{Yan2012a} for geometric treatment.

The theory for the system in pole gauge in presence of Mikhailov type reductions has been also the subject of recent research. In~\cite{GerGrahMikhVal*2012,GMV2} the case $\asl(3,\mathbb{C})$ in the presence of $\mathbb{Z}_2\times \mathbb{Z}_2$ reduction was considered (Gerdjikov--Mikhailov--Valchev system or GMV system). In~\cite{Yan2011b} it was shown that the operators found in \cite{GMV2} using classical technique and the technique developed in \cite{GKS}, are restrictions of the general position recursion operators for~$\asl(3,\mathbb{C})$, on certain subspaces of functions. In~\cite{Yan2012b} the geometric theory behind the recursion operators for the GMV system  was presented. In the present article we shall generalize the theory developed  in \cite{Yan2012a, Yan2011b,Yan2012b} for the algebra $\asl(3,\mathbb{C})$ to the algebra $\asl(n,\mathbb{C})$.  Another new feature is that we discuss here in more depth the fundamental f\/ields of the Poisson--Nijenhuis structure (P-N structure) related to the GMV system and its generalizations.

\section[Recursion operators for the GZS $\asl(n,\mathbb{C})$-system in pole gauge]{Recursion operators for the GZS $\boldsymbol{\asl(n,\mathbb{C})}$-system in pole gauge}
\subsection{Preliminary results}

Let us consider the GZS system in canonical gauge in the case $\mathfrak{g}=\asl(n,\mathbb{C})$ --  the algebra of all traceless $n\times n$ complex matrices. The
Cartan subalgebra $\mathfrak{h}$ consists of all traceless diagonal matrices and the space $\mathfrak{h}^{\perp}$ of all of\/f-diagonal matrices. The Killing form can be expressed through the trace form, we have $\langle X,Y\rangle=2n\tr XY$ for any $X,Y\in \asl(n,\mathbb{C})$, see~\cite{GoGr}. Since the element~$J$ belongs to $\mathfrak{h}$ it has the form $J=\diag(\lambda_1,\lambda_2,\ldots,\lambda_n)$, where $\sum\limits_{k=1}^n\lambda_k=0$. Next,  $J$ is regular, which means that for any $i\neq j$ we have $\lambda_i\neq \lambda_j$. In that case~$J$ `generates' the Cartan subalgebra  $\mathfrak{h}$ in the following sense. Consider the matrices
\[
J=J_1, \quad J_2=J^2-\frac{\tr J^2}{n}\mathbf{1}, \quad \ldots, \quad J_{n-1}=J^{n-1}-\frac{\tr J^{n-1}}{n}\mathbf{1}.
\]
Since $\tr J_k=0$ for $k=1,2,\ldots, n$ these matrices belong to $\mathfrak{h}$. One can easily show that these matrices are linearly independent and hence generate the Cartan subalgebra.  The same can also be deduced if one calculates the  determinant of the Gram matrix
\begin{gather}\label{eq:Grm}
G= \begin{pmatrix}
\langle J_1,J_1\rangle & \langle J_1,J_2\rangle& \dots & \langle J_1,J_{n-1}\rangle\\
\langle J_2,J_1\rangle & \langle J_2,J_2\rangle& \dots & \langle J_2,J_{n-1}\rangle\\
\dotfill&\dotfill& \dots &\dotfill\\
\langle  J_{n-1},J_1\rangle& \langle J_{n-1},J_2\rangle& \dots & \langle J_{n-1},J_{n-1}\rangle
\end{pmatrix}.
\end{gather}
Using that  $\langle X,Y\rangle=2n\tr XY$ we obtain
\[
\det G=2^{n-1}n^{n-2}\prod\limits_{i<j}(\lambda_i-\lambda_j)^2=2^{n-1}n^{n-2}D^2.
\]
The determinant of the Gram matrix~\eqref{eq:Grm} enters into the explicit calculation of the recursion operators (because we need the inverse of $G$)  and as we see it is proportional to the square of the Vandermonde determinant  $D$ constructed from the values $\lambda_1,\lambda_2,\ldots, \lambda_n$. This observation has been made for the case of the algebra $\asl(3,\mathbb{C})$, see \eqref{eq:D3case}. We see now that it is a general result for $\asl(n,\mathbb{C})$.

Since the Killing form for a given simple Lie algebra is invariant under the adjoint action of the corresponding group, we get that the Gram matrix $(\langle S_i, S_j\rangle)_{1\leq i,j\leq n-1}$ where
\[
S=S_1=\Ad\big(\psi^{-1}_0\big)J_1, \quad S_2=\Ad\big(\psi^{-1}_0\big)J_2, \quad \ldots, \quad S_{n-1}=\Ad\big(\psi^{-1}_0\big)J_{n-1}
\]
coincides with $G$. Consequently its determinant is dif\/ferent from zero.

Another observation that we want to make is that $J$ and $S$ satisfy the equation
\[
\prod\limits_{i=1}^n (X-\lambda_i\mathbf{1})=0.
\]
(Since $\lambda_i$ are the eigenfunctions of $J$ and $S$ the above is just the Cayley--Hamilton theorem.) This equation can be written in the form
\[
X^n-a_1X^{n-1}+a_2X^{n-2}+\cdots +(-1)^na_n\mathbf{1}=0,
\]
where the coef\/f\/icients $a_s$ are homogeneous polynomials of degree $s$ in $\lambda_1,\lambda_2,\ldots,\lambda_n$. If necessary, one can express them in terms the symmetric polynomials $\langle J_s,J_s\rangle$, i.e., polynomials in $\lambda_i$, $1\leq i\leq n$, using the Newton formulae but it is not needed for our purposes. We only note that $a_1=\tr J=0$.

At the end of our preparations let us denote  $S_{s;x}=\partial_x S_s$. Then we have:
\begin{proposition}\label{prop:ort}
The matrices $S_{1;x},S_{2;x},\ldots, S_{n-1;x}$ belong to $\mathfrak{h}_S^{\perp}$.
\end{proposition}
\begin{proof} We need to prove that for all $1\leq k,l\leq n-1$ we have $\langle S_k, S_{l;x}\rangle=0$. Since $S_k=S^k-(\tr J^k/n )\mathbf{1}$ we must prove
that $\tr (S^k (S^{l})_x)=0$. Using the properties of the trace we see that this is equivalent to $\tr (S^{k+l-1}S_x)=0$. On the other hand, for any integer $m>0$ we have $\tr S^m=\tr J^m=\const$. Therefore, $\partial_x\tr S^m=m\tr S^{m-1}S_x=0$ and our result follows.
\end{proof}

\subsection{Calculation of the recursion operators}

Now we pass to the calculation of the recursion operator(s) $\tilde{\Lambda}_\pm$  for the GZS  system in pole gauge on $\asl(n,\mathbb{C})$.  We shall use the equation
\[
\ri \pd{\tilde{w}},{x} - \lambda [S, \tilde{w}] = 0,
\]
which is satisf\/ied by every function of the type $\tilde{w} = \tilde{\psi} A (\tilde{\psi})^{-1}$, where $A$ is a constant matrix and~$\tilde{\psi}$ is a
fundamental solution of \eqref{eq:GZSpole}. We have
\begin{gather*}%\label{4.18}
\tilde{w} = \tilde{w}^{\rm h} + \tilde{w}^{\rm a}, \qquad \tilde{w}^{\rm h} \in \mathfrak{h}_S, \qquad \tilde{w}^{\rm a} \in {\mathfrak{h}}_S^{\perp}.
\end{gather*}
(Note that $\ker\ad_S=\mathfrak{h}_S$, the space $\mathfrak{h}_S^{\perp}$ is its orthogonal space with respect to the Killing form and by upper indices `h' and `a' we denote projections onto these spaces.) We have seen that the matrices $\{S_i\}_{i=1}^{n-1}$ (where $S_1=S$) span $\mathfrak{h}_S$. Therefore
\[
\tilde{w}^{\rm h} = \sum_{i=1}^{n-1}a_k(x)S_k,
\]
where $a_k(x)$ are coef\/f\/icient functions. We have
\[
\ri \partial_x\left[ \tilde{w}^{\rm a} + \sum_{k=1}^{n-1}a_kS_k
\right] - \lambda [ S, \tilde{w}^{\rm a}] = 0,
\]
or in other words,
\begin{gather}\label{4.19a}
\ri \partial_x\tilde{w}^{\rm a} +\ri\sum_{k=1}^{n-1}(a_k)_xS_k+\ri\sum_{k=1}^{n-1}a_kS_{k;x} - \lambda [ S, \tilde{w}^{\rm a}] = 0.
\end{gather}
In order to f\/ind the coef\/f\/icients $a_k$ we calculate the inner product of the left hand side of~\eqref{4.19a} with $S_j$. Then, taking into account Proposition~\ref{prop:ort}, we arrive at the following system
\[
\sum\limits_{k=1}^{n-1}G_{jk}(a_k)_x=-\langle \tilde{w}_x^{\rm a}, S_j\rangle=\langle \tilde{w}^{\rm a}, S_{j;x}\rangle,
\]
where $G_{ik}=\langle S_i,S_k\rangle=\langle J_i, J_k\rangle$ are the entries of the matrix $G$ introduced earlier. Equivalently,
\[
a_{kx}=(a_k)_x=\sum\limits_{k=1}^{n-1}\big(G^{-1}\big)_{ks}\langle \tilde{w}^{\rm a}, S_{s;x}\rangle.
\]
Assuming that for the eigenfunctions of $\tilde{\Lambda}_{+}$ we have $\lim\limits_{x \to
+\infty}a_s =0$ and for the eigenfunctions of~$\tilde{\Lambda}_{-}$ we have $\lim\limits_{x \to -\infty} a_s =0$  we f\/ind that
\[
a_{k}=\sum\limits_{k=1}^{n-1}\big(G^{-1}\big)_{ks}\partial_x^{-1}\langle \tilde{w}^{\rm a}, S_{s;x}\rangle,
\]
where $\partial_x^{-1}$ stands for one of the two operators
\[
\int_{+\infty}^{x} \cdot\, \rd y, \qquad \int_{-\infty}^{x} \cdot\, \rd y .
\]
Consequently,  inserting the functions $a_k$ into \eqref{4.19a}, we obtain  $\tilde{\Lambda}_\pm(\tilde{w}^a)=\lambda \tilde{w}^a$ where
\begin{gather}\label{4.24}
\tilde{\Lambda}_\pm(\tilde{Z})=\ri  \ad_S^{-1}\pi_S \left\{ \partial_x\tilde{Z} + \sum\limits_{k, s=1}^{n-1}\big(G^{-1}\big)_{ks}\partial_x^{-1}\langle \tilde{Z}, S_{s;x}\rangle S_{k;x}\right\},
\end{gather}
or equivalently,
\begin{gather}\label{4.24a}
\tilde{\Lambda}_\pm(\tilde{Z})= \ri \ad_S^{-1}\pi_S \left\{ \partial_x\tilde{Z} - \sum\limits_{k, s=1}^{n-1}\big(G^{-1}\big)_{ks}\partial_x^{-1}\langle \tilde{Z}_x, S_{s}\rangle S_{k;x}\right\}.
\end{gather}
The above operators act on functions $\tilde{Z}(x)$ that are smooth, rapidly decaying and such that $\tilde{Z}(x)\in \mathfrak{h}^{\perp}_{S(x)}$.
The formulae \eqref{4.24}, \eqref{4.24a} give us the recursion operators but they can be written in more concise form if we introduce:
\begin{itemize}\itemsep=0pt
\item The row vectors
\[
\mathbf{S}=(S_1,S_2,\ldots, S_{n-1}), \qquad \mathbf{S}_x=(S_{1;x},S_{2;x},\ldots, S_{n-1;x}).
\]
\item The column vectors
\begin{gather*}
\langle \tilde{Z}, \mathbf{S}_x \rangle=\big(\langle \tilde{Z}, S_{1;x} \rangle, \langle \tilde{Z}, S_{2;x} \rangle,\ldots, \langle \tilde{Z}, S_{n-1;x} \rangle\big)^t,  \\
\langle \tilde{Z}_x, \mathbf{S}\rangle=\big(\langle \tilde{Z}_x, S_{1} \rangle, \langle \tilde{Z}_x, S_{2} \rangle,\ldots, \langle \tilde{Z}_x, S_{n-1} \rangle\big)^t.
\end{gather*}
\end{itemize}
Then the recursion operators acquire the form
\begin{gather}\label{4.25aa}
\tilde{\Lambda}_\pm(\tilde{Z})=\ri  \ad_S^{-1}\pi_S \big\{ \partial_x\tilde{Z} +\mathbf{S}_xG^{-1}\partial_x^{-1}\langle \tilde{Z}, \mathbf{S}_{x}\rangle\big\}
\end{gather}
or
\begin{gather}\label{4.25ab}
\tilde{\Lambda}_\pm(\tilde{Z})= \ri  \ad_S^{-1}\pi_S \big\{ \partial_x\tilde{Z} - \mathbf{S}_xG^{-1} \partial_x^{-1}\langle \tilde{Z}_x, \mathbf{S}\rangle\big\}.
\end{gather}

It is easy to check now that the recursion operators for the GZS systems on the algebras $\asl(2,\mathbb{C})$, $\asl(3,\mathbb{C})$ (see~\cite{GerYan84,Yan93,Yan2011a}) are obtained from the general expressions~\eqref{4.25aa},~\eqref{4.25ab} as particular cases.

The last thing that remains to be done is to express the operator
${\ad}_S^{-1}$ through $S$. For this note that, if all the
eigenvalues of $J$ are dif\/ferent, the operators $ {\ad}_J$ and
${\ad}_J^{-1}$ considered on $\mathfrak{h}^{\perp}$ ($ {\ad}_S$ and ${\ad}_S^{-1}$ considered on $\mathfrak{h}_S^{\perp}$ respectively) are
simple and have common eigenvectors. Then we can apply the
following proposition which is actually the spectral decomposition
theorem for a given simple matrix $A$, see \cite{Lank69}.
\begin{proposition}\label{pr:sm}
Let $A$ be a simple matrix with eigenvalues $\lambda_1, \lambda_2,\ldots,
\lambda_m$ over $\mathbb{R}$ or $\mathbb{C}$. Let $\mu_1,\mu_2,\ldots,\mu_m$ be arbitrary numbers. Let us define the polynomial
\[
f(\lambda) = \sum\limits_{k=1}^m \mu_k l_k(\lambda),
\]
where $l_k$ are the Lagrange interpolation polynomials
\[
l_k(\lambda) = \prod_{i;i \neq k} \frac{(\lambda - \lambda_i)}{
(\lambda_k - \lambda_i)}.
\]
Then the matrix $f(A)$ has as eigenvalues $\mu_1, \mu_2, \ldots, \mu_m$ and the same eigenvectors as $A$ and the polynomial $f(\lambda)$ is the polynomial of minimal degree having that property.
\end{proposition}
\begin{remark} It is not dif\/f\/icult to see that $l_k(A)$ is the projector onto the subspace corresponding to the eigenvalue $\lambda_k$ in the splitting of the space into eigen\-spaces of the matrix $A$.
\end{remark}
\begin{remark}
In case the matrix $A$ is not simple one again can produce a polynomial  $f(\lambda)$ of minimal degree having the property stated in the Proposition \ref{pr:sm} though its construction is more complicated, see \cite{Lank69}. In case just $\ad_J^{-1}$ is needed, one can also use the following procedure:
\begin{enumerate}\itemsep=0pt
\item The minimal polynomial $m(\lambda)$ of $\ad_J$ (on the whole algebra) is a product $m(\lambda)=\lambda m_1(\lambda)$ and $\lambda$ and
$m_1(\lambda)$ are co-prime.
\item The algebra splits into direct sum of invariant subspaces $\ker\ad_J$ and $\im \ad_J$ (because $\ad_J$ is skew-symmetric) and the minimal polynomials of  $\ad_J$ on these spaces are $\lambda$ and $m_1(\lambda)$ respectively.
\item  On $\im \ad_J$ the operator $\ad_J$ is invertible and one can f\/ind $\ad_J^{-1}$ as polynomial in $\ad_J$ multiplying the equation $m_1(\ad_J)=0$  by $\ad_J^{-1}$.
\end{enumerate}
Note that the polynomial $g(\lambda)$ such that $\ad_J^{-1}=g(\ad_J)$ will have now degree $\deg(m_1)-1$.
\end{remark}
In the case of an arbitrary semisimple Lie algebra, assuming that all the values $\alpha(J)$, $\alpha\in \Delta$ are dif\/ferent, the operators $\ad_J$ and ${\ad}_J^{-1}$  have  eigenvalues $\alpha(J)$ and $1 /\alpha(J)$, $\alpha \in \Delta$ respectively and eigenvectors $E_\alpha$, $\alpha \in \Delta$. Then ${\ad}_J^{-1}$ is equal to $l({\ad}_J)$, where $l(\lambda)$ is the polynomial
\begin{gather}\label{eq:invad1}
l(\lambda) = \sum\limits_{\alpha \in \Delta_{+}} \frac{\lambda r_\alpha (\lambda)}{\alpha(J)^2 },\qquad
r_\alpha(\lambda) =  \prod_{\beta \in \Delta_{+},\,  \beta \neq \alpha}
\frac{\lambda^2 - \beta^2(J)}{\alpha(J)^2 - \beta(J)^2}.
\end{gather}
In our case the roots are $\alpha_{ij} = \varepsilon_i - \varepsilon_j$, $i\neq j$ where on diagonal matrices $A$ with diagonal  elements~$a_i$ we have $\varepsilon_i(A)=a_i$.  So $\alpha_{ij}(J) = \lambda_i - \lambda_j$ and the expression for $\ad_S^{-1}$ can be written easily, we shall do it a little further. What we obtained already allows us to make an important observation. As the polynomial $l(\lambda)$ from \eqref{eq:invad1} is of the form $\lambda l_0(\lambda)$, where $l_0$ is another polynomial, then we have $l({\ad}_J) = l_0({\ad}_J) \ad_J$. Therefore we have $l({\ad}_J)\pi_0 = l({\ad}_J)$. In the same way
\begin{gather}\label{4.30}
l({\ad}_S) =  \ad_S^{-1}, \qquad l({\ad}_S){\pi}_S = l({\ad}_S), \qquad \pi_S=\ad_J^{-1}\ad_S.
\end{gather}
Therefore, if one assumes that $\ad_S^{-1}$ is given by a polynomial in $\ad_S$, writing the projector $\pi_S$ in the expression for the recursion operators~\eqref{4.25aa},~\eqref{4.25ab} is redundant.

Continuing our discussion about $\ad_S^{-1}$,  as we already remarked, in the case when we need only to express $\ad_S^{-1}$ through $\ad_S$,  instead of Proposition \ref{pr:sm} we can use the minimal polynomial for $\ad_S$  (restricted to $\mathfrak{h}^{\perp}$ of course).  Indeed, in the case of an arbitrary semisimple Lie algebra $\mathfrak{g}$ and regular $J$ let us assume that all the values  $\alpha(J)$ are dif\/ferent. Then the minimal polynomial for $\ad_J$ on $\mathfrak{h}^{\perp}$ has the form
\[
m(\lambda)=\prod\limits _{\alpha\in \Delta_{+}}\big(\lambda^2-\alpha(J)^2\big)=\sum\limits_{k=0}^{p}a_k\lambda^{2k}, \qquad a_p=1.
\]
Here $p=\frac 12 \dim(\mathfrak{h}^{\perp})=\frac 12 (\dim\mathfrak{g}-\dim \mathfrak{h})$ and,  because $J$ is regular,
\[
a_0=(-1)^p\prod_{\alpha\in \Delta_{+}}\alpha(J)^2\neq 0.
\]
Since $m(\ad_S)=0$ we obtain
\begin{gather}\label{eq:invad2}
\ad_S^{-1}=\ad_SR(\ad_S), \qquad \pi_S=\ad_S^{-1}\ad_S=\ad_S R(\ad_S)\ad_S,
\end{gather}
where $R(\lambda)$ is the polynomial
\begin{gather*}
R(\lambda)=-\frac{1}{a_0}\big(a_1\lambda^2 +a_2\lambda^4+\dots+a_p\lambda^{2p}\big).
\end{gather*}
Both expressions \eqref{4.30} and \eqref{eq:invad2} for $\ad_S^{-1}$ give the same result in the case where all $\alpha(J)$ are dif\/ferent.  Indeed, both polynomials $\lambda l(\lambda)$ and $\lambda R(\lambda)$ are monic and when $\lambda=\alpha(J)$, $\alpha\in \Delta$ give~$1/\alpha(J)$. Since they are of degree $2p-1$   these polynomials coincide. In particular, in the case of the algebra $\asl(n,\mathbb{C})$ we get
\begin{gather*}
R(\lambda)=\frac{(-1)^{n(n+1)}}{\lambda^2D^2}\left\{\prod\limits_{i<j}\big[\lambda^2-(\lambda_i-\lambda_j)^2\big]-(-1)^{n(n+1)}D^2\right\}.
\end{gather*}
Unfortunately, the expressions for $\ad_S^{-1}$ become very complicated for big $n$, hampering the possible applications.  A simplif\/ication can be obtained for some particular choices of $J$, for which the minimal polynomial of $\ad_J$ on $\mathfrak{h}^{\perp}$ has smaller degree than in the case of general~$J$.  Here are the expressions for $\ad_S^{-1}$ used up to now in the literature:
\begin{enumerate}
\item Classical Zakharov--Shabat system, $\mathfrak{g}=\asl(2,\mathbb{C})$, $J=\diag(1,-1)$ \cite{GerYan84,GerYan85}
\[
\ad_S^{-1}=\frac{1}{4}\ad_S.
\]
\item GZS system on $\mathfrak{g}=\asl(3,\mathbb{C})$ in general position, $J=\diag(\lambda_1,\lambda_2,\lambda_3)$, $\lambda_i\neq 0$. The last condition ensures that all $\alpha(J)$'s are dif\/ferent \cite{Yan93,Yan2011a}
\begin{gather}\label{eq:D3case}
\ad_S^{-1}=l(\ad_S),\qquad
l(\lambda)= \frac{\lambda}{D^2} \left( \lambda^2 -\frac{3}{2}C_2 \right)^2,
\end{gather}
where
\[
C_2=\lambda_1^2+\lambda_2^2+\lambda_3^2, \qquad D=( \lambda_1 - \lambda_2 )( \lambda_2 - \lambda_3)
( \lambda_1 - \lambda_3 ).
\]
\item GZS type system on $\mathfrak{g}=\asl(3,\mathbb{C})$ with $\mathbb{Z}_2\times \mathbb{Z}_2$ reduction (GMV system), $J=\diag(-1,0,1)$ \cite{GMV2,Yan2011a}
\begin{gather*}
\ad_S^{-1}=l(\ad_S),\qquad
l(\lambda)= -\frac{1}{4}\big( \lambda^3 -5\lambda\big).
\end{gather*}
In this case the fact that some of the eigenvalues of $\ad_J$ on $\mathfrak{h}^{\perp}$ (and consequently of~$\ad_S$ on~$\mathfrak{h}_S^{\perp}$)  are not simple leads to a decrease in the degree of the polynomial~$l(\lambda)$.
\end{enumerate}

\section{Geometric interpretation}

Fixing the element $J$ for the GZS $\mathfrak{g}$-system in pole gauge, the smooth function~$S(x)$ with domain~$\mathbb{R}$, see \eqref{eq:GZSpole}, is not subject to any restrictions except that~$S(x)\in \mathcal{O}_J$ and~$S(x)$ tends fast enough to some constant values when $x\mapsto \pm\infty$.  First let us consider the even more general case when~$S(x)$ is smooth, takes values in $\mathfrak{g}$ and when  $x\to \pm\infty$ tends fast enough to constant values. The functions of this type form an inf\/inite-dimensional manifold which we shall denote by~$\mathcal{M}$. Then it is reasonable to assume that the tangent space $T_S(\mathcal{M})$ at $S$ consists of all the smooth functions $X:\mathbb{R}\mapsto \mathfrak{g}$ that tend to zero fast enough when $x\mapsto \pm\infty$. We denote that space by~$\mathfrak{F}(\mathfrak{g})$. We shall also assume that the `dual space' $T_S^*(\mathcal{M})$ is equal to $\mathfrak{F}(\mathfrak{g})$ and if $\alpha\in T_S^*(\mathcal{M})$, $X\in T_S(\mathcal{M})$ then
\[
\alpha(X)=\langle \langle \alpha,X\rangle\rangle\equiv \int_{-\infty}^{+\infty}\langle \alpha(x),X(x)\rangle \rd x,
\]
where $\langle~,~\rangle$ is the Killing form of $\mathfrak{g}$.
\begin{remark}
In other words, we identify $T_S^*(\mathcal{M})$ and $T_S(\mathcal{M})$ using the bi-linear form $\langle\langle~,~\rangle\rangle$.
We do not want to make the def\/initions more precise, since we speak rather about geometric picture than about precise results. Such results could be obtained after a profound study of the spectral theory of $L$ and $\tilde{L}$. In particular, we have put the dual space in quotation marks because it is clearly not equal to the space of continuous linear functionals on $\mathfrak{F}(\mathfrak{g})$. We also emphasize that when we speak about `allowed' functionals $H$ on ${\mathcal M}$ we mean that $\frac{\delta H}{\delta S}\in T_S^*(\mathcal{M})\sim \mathfrak{F}(\mathfrak{g})$.
\end{remark}
Now we want to introduce some facts. The f\/irst fact is that since we identify  $T_S^*(\mathcal{M})$ and $T_S(\mathcal{M})$ the operators
\begin{gather}\label{eq:P}
\alpha\mapsto P(X)=\ri\partial_x\alpha,\qquad
\alpha\mapsto Q(\alpha)=\ad_S(\alpha),\qquad %\label{eq:Q}\\
S\in \mathcal{M}, \qquad\alpha\in T_S^*(\mathcal{M})
\end{gather}
can be interpreted as Poisson tensors on the manifold $\mathcal{M}$. This is well known, see for example~\cite{GerViYa2008}, where the  issue has been discussed in detail and the relevant references are given. One can also verify directly that if $H_1$, $H_2$ are two functions (allowed functionals) on the manifold of potentials~$\mathcal{M}$ then
\begin{gather*}
\{H_1,H_2\}_P=\left\langle\left\langle \frac{\delta H_1}{\delta S}, \partial_x\frac{\delta H_2}{\delta S}\right\rangle\right\rangle, \qquad
\{H_1,H_2\}_Q=\left\langle\left\langle \frac{\delta H_1}{\delta S}, \left[S, \frac{\delta H_2}{\delta S}\right]\right\rangle\right\rangle
\end{gather*}
are Poisson brackets. It is also known from the general theory that these Poisson tensors are compatible~\cite[Chapter~15]{GerViYa2008}. In other words $P+Q$ is also a Poisson tensor.  Note that the tensor~$Q$ is the canonical Kirillov tensor which acquires this form because the algebra is simple and consequently the coadjoint and adjoint representations are equivalent.

Now let $\mathcal{O}_J$ be the orbit of the adjoint action of $G$ passing through $J$. Let us consider the set of smooth functions $f:\mathbb{R}\mapsto \mathcal{O}_J$ such that when $x\to \pm\infty$ they tend fast enough to constant values. The set of these functions is denoted by $\mathcal{N}$ and  clearly can be considered as a~submanifold of $\mathcal{M}$. If $S\in \mathcal{N}$  the tangent space
$T_S(\mathcal{N})$ consists of all smooth functions $X$ that vanish fast enough when $x\mapsto \pm\infty$ and are such that  $X(x)\in T_{S(x)}(\mathcal{O}_J)$. (Recall that $\mathcal{O}_J$ is a~smooth manifold in the classical sense). We again assume that $T_S^*(\mathcal{N})\sim T_S(\mathcal{N})$ and that these spaces are identif\/ied via $\langle\langle~,~\rangle\rangle$.

We can try now to restrict the Poisson tensors $P$ and $Q$ from the manifold $\mathcal{M}$ to the mani\-fold $\mathcal{N}$. The problem how to restrict a Poisson tensor on a submanifold  has been  solved in principle~\cite{MaRa*86}, see also
\cite{OrtRatiu*04a,OrtRatiu*04b}. We shall use a simplif\/ied version of these results (see~\cite{MagMor*84,MagMorRagn*85}) and we shall call it the f\/irst restriction theorem:
 \begin{theorem}\label{Th:RT} Let $\mathcal{M}$ be a Poisson manifold with Poisson tensor~$P$
and $\bar{\mathcal{M}} \subset \mathcal{M}$ be a submanifold. Let us
denote by $\rj$ the inclusion map of $\bar{\mathcal{M}}$ into $\mathcal{M}$,
by $\mathcal{X}_P^*(\bar{\mathcal{M}})_m$ the subspace of covectors
$\alpha \in T^*_m(\mathcal{M})$ such that
\[
P_m(\alpha)\in \rd\rj_m(T_m(\bar{\mathcal{M}}))=\im (\rd\rj_m),\qquad m\in \bar{\mathcal{M}}
\]
$($where $\im$ denotes the image$)$, and by $T^{\perp}(\bar{\mathcal{M}})_m$
the set of all covectors at $m\in\mathcal{M}$ vanishing on the
subspace $\im(\rd \rj_m)$, $m\in \bar{\mathcal{M}}$  $($also called the
annihilator of $\im(\rd \rj_m)$ in $T^*_m(\mathcal{M}))$. Let the
following relations hold:
\begin{gather*}
\mathcal{X}_P^*(\bar{\mathcal{M}})_m+T^{\perp}(\bar{\mathcal{M}})_m=T_m^*(\mathcal{
M}),\qquad m\in \bar{\mathcal{M}},\\ %\label{eq:I}\\
\mathcal{X}_P^*(\bar{\mathcal{M}})_m\cap T^{\perp}(\bar{\mathcal{M}})_m\subset \ker(P_m),\qquad m\in
\bar{\mathcal{M}}.%\label{eq:II}
\end{gather*}
Then there exists unique Poisson tensor $\bar{P}$ on $\bar{\mathcal{M}}$, $\rj$-related with $P$, that is
\[
P_m=\rd\rj_m\circ\bar{P}_m\circ (\rd\rj_m)^*, \qquad m\in\bar{\mathcal{M}} .
\]
\end{theorem}
The proof of the theorem is constructive. One takes $\beta\in T_m ^*(\bar{\mathcal{M}})$, then represents  $(\rj^*\beta)_m$ as $\alpha_1+\alpha_2$ where $\alpha_1\in \mathcal{X}_P^*(\bar{\mathcal{M}})_m$, $\alpha_2\in  T^{\perp}(\bar{\mathcal{M}})_m$ and puts $\bar{P}_m(\beta)=P_m(\alpha_1)$ (we identify $m$ and~$\rj(m)$ here).

The restriction we present below has been carried out in various works in the simplest case $\mathfrak{g}=\asl(2,\mathbb{C})$, see for example \cite{MagMorRagn*85}. We do it now in the case $\mathfrak{g}=\asl(n,\mathbb{C})$. Restricting the Poisson tensor $Q$ is easy, one readily gets that the restriction $\bar{Q}$ is given by the same formula as before
\begin{gather}\label{eq:barQ}
\alpha\mapsto \bar{Q}(\alpha)=\ad_S(\alpha), \qquad S\in \mathcal{N},\qquad  \alpha\in T_S^*(\mathcal{N}).
\end{gather}
The tensor $P$ is  a little harder to restrict. Let us introduce some notation and facts f\/irst. Since $J$ is a regular element from the Cartan subalgebra $\mathfrak{h}$, each element $S$ from the orbit $\mathcal{O}_J$  is regular. Therefore $\mathfrak{h}_S\equiv \ker\ad_{S}$ is a Cartan subalgebra of $\asl(n,\mathbb{C})$ and we have
\begin{gather*}%\label{eq:splitt}
\asl(n,\mathbb{C})=\mathfrak{h}_S (x)\oplus \mathfrak{h}_S^{\perp}(x).
\end{gather*}
If $X\in T_S(\mathcal{N})$ then $X(x)\in \mathfrak{h}_S^{\perp}(x)$ and $X$ vanishes rapidly when $x\mapsto \pm \infty$. We shall denote the set of these functions by $\mathfrak{F}(\mathfrak{h}_S^{\perp})$ so $X\in \mathfrak{F}(\mathfrak{h}_S^{\perp})$ (this means a little more than simply $X\in\mathfrak{h}_S^{\perp}$). Using the same logic, for $X\in \mathfrak{F}(\mathfrak{h}_S^{\perp})$ we write $\ad_S(X)\in \mathfrak{F}(\mathfrak{h}_S^{\perp})$ which means that the function $\ad_{S(x)}X(x)$ belongs to $\mathfrak{F}(\mathfrak{h}_S^{\perp})$.
Now we are in a position to perform the restriction of $P$ on $\mathcal{N}$. For $S\in \mathcal{N}$ we have
\begin{gather*}
\mathcal{X}^*_P(\mathcal{N})_S=\{\alpha:\ri\partial_x\alpha\in \mathfrak{F}(\mathfrak{h}_S^{\perp})\},\qquad
T^{\perp}(\mathcal{N})_S=\{\alpha:\langle\langle \alpha,X\rangle\rangle=0, ~X\in \mathfrak{F}(\mathfrak{h}_S^{\perp})\}.
\end{gather*}
We see that $T^{\perp}(\mathcal{N})_S$ is the set of smooth functions $\alpha$ such that $\alpha\in \mathfrak{h}_S$ and such that they vanish fast enough when $x\mapsto\pm \infty$. We shall denote this space by
$\mathfrak{F}(\mathfrak{h}_S)$. We introduce also the space of functions $X\in \mathfrak{h}_S$ which tend rapidly to some constant values when $x\mapsto \pm\infty$ and denote this space by $\mathfrak{F}(\mathfrak{h}_S)_{0}$. Clearly,  since $S_k$, $k=1,2,\ldots, n-1$ span $\mathfrak{h}_S$, we have
\begin{gather*}
\mathfrak{F}(\mathfrak{h}_S)_0= \left\{ X:X= \sum\limits_{k=1}^{n-1}a_k(x)S_k(x), \  a_k(x) \ \mbox{smooth},\
a_k(x) \  \mbox{tend to $\const$ as} \ x\mapsto\pm\infty\right\},  \\
\mathfrak{F}(\mathfrak{h}_S)=\left\{X:X= \sum\limits_{k=1}^{n-1}a_k(x)S_k(x), \ a_k(x) \ \mbox{smooth}, \lim_{x\to\pm\infty}a_k(x)=0\right\}.\nonumber
\end{gather*}
In what follows we shall adopt matrix notation and shall denote by $A(x)$ the column with components $a_k(x)$. Then $\mathcal{X}^*_P(\mathcal{N})_S \cap T^{\perp}(\mathcal{N})_S$ consists of the elements
\[
\alpha=\sum\limits_{k=1}^{n-1}a_k(x)S_k(x)=\mathbf{S}A,
\]
 with $a_k(x)$  vanishing at inf\/inity and such that $\ri\partial_x\alpha\in \mathfrak{F}(\mathfrak{h}_S^{\perp})$.
But
\[
\ri\partial_x\alpha=\ri\mathbf{S}_xA+\ri\mathbf{S}A_x,
\]
so we must have $A_x=0$ and $a_s(x)$  are constants. But $a_s(x)$  must also vanish at inf\/inity so we see that $a_s=0$. Consequently $\mathcal{X}^*_P(\mathcal{N})_S \cap T^{\perp}(\mathcal{N})_S=\{0\}\subset \ker P_S$.

Let us take now arbitrary $\alpha\in T^{*}(\mathcal{N})_S $. We want to represent it as $\alpha_1+\alpha_2$, $\alpha_1\in \mathcal{X}^{*}(\mathcal{N})_S$, $\alpha_2\in T^{\perp}(\mathcal{N})_S$. First of all,
$\alpha_2=\mathbf{S}B(x)$ where $B(x)$ a column with components $b_s(x)$~-- scalar functions that vanish at inf\/inity. This means that
\[
\ri\partial_x\alpha=\ri\partial_x\alpha_1+\ri\mathbf{S}_xB+\ri\mathbf{S}B_x,
\]
where $\ri\partial_x\alpha_1\in \mathfrak{F}(\mathfrak{h}_S^{\perp})$. But then we have
\[
\langle \partial_x\alpha, \mathbf{S}(x)\rangle=GB_x,
\]
where $G$ is the Gram matrix we introduced earlier. Therefore
\[
B=G^{-1}\partial_x^{-1}\langle \partial_x\alpha(x), \mathbf{S}(x)\rangle .
\]
\begin{remark}
In the theory of recursion operators when one calculates the hierarchies of NLEEs or the conservation laws the expressions on which the operator $\partial_x^{-1}$ acts  are total derivatives. Thus the same results will be obtained choosing for $\partial_x^{-1}$ any of the following operators
\[
\int_{-\infty}^x \cdot \, \rd y,\qquad \int_{+\infty}^x \cdot\, \rd y,\qquad \frac{1}{2}\left(\int_{-\infty}^x \cdot \, \rd y+\int_{+\infty}^x\cdot\,  \rd y\right).
\]
However, more frequently one uses the third expression when one writes the corresponding Poisson tensors in order to make them explicitly skew-symmetric.
\end{remark}
Returning to our task, let us put
\begin{gather*}
\alpha=\alpha_1+\alpha_2,\qquad
\alpha_1=\alpha-\alpha_2,\qquad
\alpha_2=\mathbf{S}G^{-1}\partial_x^{-1}\langle \partial_x\alpha(x), \mathbf{S}(x)\rangle ,
\end{gather*}
where $\alpha_1$ and $\alpha_2$ lie in the spaces $\mathcal{X}^{*}(\mathcal{N})_S$ and $T^{\perp}(\mathcal{N})_S=\mathfrak{F}(\mathfrak{h}_S)$ respectively. We note also that $\rd\rj_S^*\beta=\pi_S(\beta)$.
Thus the conditions of  the f\/irst restriction theorem are fulf\/illed and if $\beta \in T^*_S(\mathcal{N})$ the restriction $\bar{P}$ of $P$ on $\mathcal{N}$ has the form
\begin{gather*}
\bar{P}(\beta)=\ri\pi_S\partial_x\beta-\ri\mathbf{S}_x\partial_x^{-1}\langle \partial_x\beta(x), \mathbf{S}(x)\rangle.
\end{gather*}
The tensor $\bar{Q}$ is invertible on $\mathcal{N}$, so one can construct a Nijenhuis tensor $N=\bar{P}\circ \ad_S^{-1}$
\begin{gather}\label{eq:NGP1}
N(X)=\ri\pi_S\partial_x(\ad_S^{-1}X )-\ri\mathbf{S}_xG^{-1}\partial_x^{-1}\langle \partial_x(\ad_S^{-1}X), \mathbf{S}(x)\rangle, \qquad X\in \mathfrak{F}(\mathfrak{h}_S^{\perp}).
\end{gather}
Taking into account that $\langle \ad^{-1}_S(X), \mathbf{S}\rangle=0$,  the above can be cast into the equivalent form
\begin{gather}\label{eq:NGP2}
N(X)=\ri\pi_S\partial_x(\ad_S^{-1}X )+\ri\mathbf{S}_xG^{-1}\partial_x^{-1}\langle\ad_S^{-1}X, \mathbf{S}_x\rangle, \qquad X\in \mathfrak{F}(\mathfrak{h}_S^{\perp}).
\end{gather}
From the general theory of compatible Poisson tensors it now follows that
\begin{theorem}
The Poisson tensor field $\bar{Q}$ \eqref{eq:barQ} and the Nijenhuis tensor field $N$  given by~\eqref{eq:NGP1},~\eqref{eq:NGP2} endow the manifold $\mathcal{N}$ with a P-N structure.
\end{theorem}

The f\/inal step is to calculate the dual of the tensor $N$ with respect to the pairing $\langle\langle~,~\rangle\rangle$. A~quick calculation, taking into account that $\ad_S$ is skew-symmetric with respect to the Killing form, gives
\[
N^*(\alpha)=\ri\ad_S^{-1}\left[\pi_S\partial_x\alpha +\mathbf{S}_xG^{-1}\langle\alpha, \mathbf{S}_x\rangle \right], \qquad \alpha\in \mathfrak{F}(\mathfrak{h}_S^{\perp}),
\]
or equivalently,
\[
N^*(\alpha)=\ri\ad_S^{-1}\left[\pi_S\partial_x\alpha -\mathbf{S}_x G^{-1}\partial_x^{-1}\langle\alpha_x,\mathbf{S}\rangle\right], \qquad \alpha\in \mathfrak{F}(\mathfrak{h}_S^{\perp}).
\]
But these are the recursion operators $\tilde{\Lambda}_{\pm}$ for the GZS system in pole gauge from~\eqref{4.25aa},~\eqref{4.25ab} and our results conf\/irm the general fact that the recursion operators and the Nijenhuis tensors are dual objects.

Now, according to the general theory of recursion operators, see \cite{GerViYa2008}, the NLEEs related to the system $\tilde{L}$ have the form
\begin{gather}\label{eq:ZSpole3}
-\ri\ad_{S}^{-1}\frac{\partial S}{\partial t}+
(\tilde{\Lambda}_{\pm})^n\tilde{\Lambda}_{\pm} \pi_S H=0,
\end{gather}
where  $H$  is an element of the Cartan subalgebra of $\asl(n,\mathbb{C})$. As discussed earlier, $\pi_S$ can be expressed as a polynomial in $\ad_S$. Then \eqref{eq:ZSpole3} gives the hierarchies of NLEEs related to the GZS system in pole gauge as hierarchies of equations gauge-equivalent to hierarchies related to the GZS system in canonical gauge. If, however, one is not interested in f\/inding the pairs of gauge-equivalent equations, but wants just to f\/ind the NLEEs, one can proceed as follows. Recall that recursion operators also produce the hierarchy of Lax pairs. In fact, if a NLEE has Lax representation $[\tilde{L},\tilde{A}]=0$ with $\tilde{L}$ in general position and $\tilde{A}$ has the form
\[
\tilde{A}=\ri\partial_t+\sum\limits_{k=0}^n\lambda^k\tilde{A}_k, \qquad \tilde{A}_k\in \mathfrak{h}_S,
\]
then $\ri\partial_x A_k \in \mathfrak{h}_S^{\perp}$,  $A_0=\const$. Using a gauge transformation depending only on $t$ one can ensure that $A_0=0$ and then the coef\/f\/icients $\tilde{A}_k$ for $k=1,2,\ldots, n-1$ may be calculated with the help of the recursion operator in the following way
\begin{gather*}
\tilde{A}_{k-1}^{\rm{a}}=\tilde{\Lambda}_{\pm}\tilde{A}_{k}^{\rm{a}},\qquad
\tilde{A}_{k}^{\rm{d}}=\ri\mathbf{S}G^{-1}\partial_x^{-1}\langle \tilde{A}_{k}^{\rm{a}},\mathbf{S}_x\rangle, \qquad
\tilde{A}_s^{\rm{d}}=(\id-\pi_S)\tilde{A}_s, \qquad \tilde{A}_s^{\rm{a}}=\pi_S \tilde{A}_s  .
\end{gather*}
For the coef\/f\/icient $\tilde{A}_{n-1}$ one has that $\ri\partial_x\tilde{A}_n=[S,\tilde{A}_{n-1}]$ and, since $\tilde{A}_n\in \mathfrak{h}_S$, there are $n-1$ scalar functions $a_k(x)$ forming a column vector $a(x)$ such that $A_n=\mathbf{S}(x)a(x)$.
This gives
\[
\ri\mathbf{S}_x a(x)+\ri\mathbf{S}a_x(x)\in \mathfrak{h}_S^{\perp},
\]
and therefore $a_x=0$ so $a_k(x)$ are constants. Thus $\tilde{A}_{n-1}=\ri\ad_S^{-1}(\mathbf{S}a)$
 and the hierarchy of the NLEEs related to $\tilde{L}$  is
\begin{gather*}%\label{eq:EEZSSspole3a}
-\ad_{S}^{-1}\frac{\partial S}{\partial t}+(\tilde{\Lambda}_{\pm})^n \big(\ad_S^{-1}(\mathbf{S}_xa)\big)=0, \qquad a=(a_1,a_2,\ldots,a_{n-1})^t.
\end{gather*}

\section{Reductions}
\subsection{Algebraic aspects}
As a matter of fact the situation most interesting for applications is when we do not have the GZS system in pole gauge in the general case, but when additional restrictions are made. For example, in the Heisenberg ferromagnet equation  we require that $S=S^{\dag}$ (where $\dag$ stands for Hermitian conjugation). Similarly, for systems describing spin chain dynamics that we mentioned in the case of $\asl(3,\mathbb{C})$, we also require $S^{\dag}=S$. All this means that we require for $n=2,3$ that $S$ belongs to $\ri\su(n)$. The algebra $\su(n)$ is a real form of $\asl(n,\mathbb{C})$ with respect to complex conjugation $\sigma(X)=-X^{\dag}$ and
\begin{gather*}
\asl(n,\mathbb{C})=\su(n)\oplus \ri \su(n),\qquad [\su(n),\su(n)]\subset \su(n),  \\
[\su(n),\ri\su(n)]\subset \ri\su(n),\qquad [\ri\su(n),\ri\su(n)]\subset \su(n).
\end{gather*}
The space $\ri\su(n)$ is the space of $n\times n$ Hermitian matrices. Of course, it is not a real Lie algebra since, if $X,Y\in \ri\su(n)$, then $[X,Y]\in \su(n)$  and hence $\ri[X,Y]\in \ri \su(n)$.

Next, since the Cartan subalgebra $\mathfrak{h}$ is invariant under $\sigma$, it also splits
\[
\mathfrak{h}=(\mathfrak{h}\cap \su(n))\oplus (\mathfrak{h}\cap \ri\su(n)).
\]
The f\/irst space  consists of diagonal matrices with purely imaginary entries, while the second consists of diagonal matrices with real entries. If we want $S$ to be Hermitian, then it is natural to assume that $J$ is real.  In fact the reduction one obtains in this way is most ef\/fectively treated by the notion of Mikhailov's reduction group, see \cite{Mikh3,Mikh1,Mikh2}. According to that concept, in order to perform a reduction we must have a group $G_0$ acting on the fundamental solutions of the linear problems $L$ and $A$ in the Lax representation. Assume we have the GZS system in pole gauge and we take the group generated by one element $g_1$, acting on the fundamental solutions as
\begin{gather}\label{eq:g1}
g_1(\tilde{\psi}(x;\lambda)) = [\tilde{\psi}^{-1}(x;\lambda^*)]^{\dag},
\end{gather}
where $^*$ stands for complex conjugation. Since $g_1^2=\id$, the group generated by $g_1$ is isomorphic to $\mathbb{Z}_2$ and we have $\mathbb{Z}_2$-reduction. The invariance of the set of fundamental solutions under the action \eqref{eq:g1} means that $\sigma(S)=-S$ (that is $S\in \ri\su(n)$) and that $S$ belongs to the orbit of $\SU(n)$ passing through $J\in \ri\su(n)$. Let us denote this orbit by $\mathcal{O}_J(\SU(n))$. Thus $J$ must be real.
Moreover, if the operator $A$ has the form
\[
A=\ri\partial_t+\sum\limits_{k=0}^NA_k\lambda^k,
\]
then, if the set of fundamental common solutions $\tilde{\psi}$ of \eqref{eq:GZSpole}  and $A\tilde{\psi}=0$ is $G_0$-invariant, one must have also
$\sigma(A_k)=A_k$ where $\sigma$ is the complex conjugation introduced earlier, that is $A_k\in \ri\su(n)$.  One may note that the generating operators are the same as before simply we may assume that all matrices are in $\ri\su(n)$.

One can have also another complex conjugation of $\asl(n,\mathbb{C})$. Denote it by $\tau$. Then one can construct another $\mathbb{Z}_2$ reduction. One can, for example, take $\tau(X)=X^*$ and
$g$ acting as
\[
g(\tilde{\psi}(x;\lambda))=(\tilde{\psi}(x;\lambda^*)^*.
\]
Then one obtains that $S^*=-S$, $A_k^*=-A_k^*$, so introducing $S^0=\ri S$ and $A^0_k=\ri A_k$ one sees that (after canceling $\ri$) the $L$, $A$ pair is real, with matrices belonging to $\asl(n,\mathbb{R})$.

It is possible to have restrictions def\/ined by two complex conjugations. This can be achieved in the following way. Suppose $\tau$ is another complex conjugation, commuting with $\sigma$. Then as it is well known $h=\sigma\tau=\tau\sigma$ is an involutive automorphism of the algebra $\asl(n,\mathbb{C})$. In fact doing the things the other way round is easier -- to f\/ind an involutive automorphism $h$  commuting with $\sigma$ and then put $\tau=h\sigma=\sigma h$. One can take, for example, $h(X)=J_KXJ_K$, where $J_K$ is a~diagonal matrix
\[
\diag(\underbrace{1,1,\ldots, 1}_{K \ \text{times}}, \underbrace{-1,-1,\ldots, -1}_{n-K \ \text{times}}),
\]
and one can put $\tau=h\sigma=\sigma h$.  Then one can consider the group $G_0$ with generating elements~$g_1$ (as in~\eqref{eq:g1}) and $g_2$, where
\begin{gather}\label{eq:g2}
g_2(\tilde{\psi}(x;\lambda))=h(\tilde{\psi}(x;-\lambda))=J_K(\tilde{\psi}(x;-\lambda))J_K.
\end{gather}
Since $g_1g_2=g_2g_1$, $g_1^2=g_2^2=\id$ the reduction group $G_0$ is isomorphic to $\mathbb{Z}_2\times\mathbb{Z}_2$ so the invariance of the fundamental solutions with respect to the actions~\eqref{eq:g1} and~\eqref{eq:g2} def\/ines  $\mathbb{Z}_2\times\mathbb{Z}_2$ reduction.  In~\cite{GMV2, GMV1} the above reduction has been applied in the case of the algebra $\asl(3,\mathbb{C})$ with $J_1=\diag(1,-1,-1)$ (here $K=1$). A similar reduction group but this time  related to the algebra $\so(5)$ has been considered recently in~\cite{Valchev*2011}.

Returning to the general case, note that  instead of the generators $g_1$, $g_2$ we could use $g_1$, $g_1g_2$ and
\[
g_1g_2(\tilde{\psi}(x;\lambda))=J_K\big[\tilde{\psi}^{-1}(x;-\lambda^*))\big]^{\dag}J_K.
\]
The invariance with respect to the group generated by $g_1$, $g_2$ means that
\begin{gather*}
\sigma(S)=-S, \qquad\! h(S)=-S, \qquad\!\! \sigma(A_{s})=-A_s, \qquad\!\! h(A_{2k})=A_{2k}, \qquad\!\! h(A_{2k-1})=-A_{2k-1},
\end{gather*}
or equivalently
\begin{gather*}
\sigma(S)=-S, \qquad\! \tau(S)=S, \qquad  \sigma(A_{s})=-A_s, \qquad  \tau(A_{2k})=-A_{2k}, \qquad  \tau(A_{2k-1})=A_{2k-1}.
\end{gather*}
The complex conjugation $\tau$ also splits the algebra $\asl(n,\mathbb{C})$
\begin{gather*}
\asl(n,\mathbb{C})=\mathfrak{s}\oplus \ri \mathfrak{s},\qquad
[\mathfrak{s},\mathfrak{s}]\subset \mathfrak{s}, \qquad [\mathfrak{s},\ri\mathfrak{s}]\subset \ri\mathfrak{s}, \qquad [\ri\mathfrak{s},\ri\mathfrak{s}]\subset \mathfrak{s}.
\end{gather*}
(On $\mathfrak{s}$ the map $\tau$ is equal to $\id$, on $\ri\mathfrak{s}$ to $-\id$.)
We prefer to work with the involutive automor\-phism~$h$ instead of $\tau$. Then the algebra splits into two invariant subspaces for $h$ and these spaces are orthogonal with respect to the Killing form. We have
\begin{gather*}
\mathfrak{f}_{\epsilon}=\{X: h(X)=(-1)^{\epsilon}X\}, \quad \epsilon=0,\, 1, \qquad
\asl(n,\mathbb{C})=\mathfrak{f}_0\oplus\mathfrak{f}_1.
\end{gather*}
A calculation shows that the subalgebra $\mathfrak{f}_0$ consists of matrices $U_0$ having block form
\[
U_0= \begin{pmatrix} A&0\\0&B\end{pmatrix}, \qquad \mbox{where} \quad \tr A+\tr B=0,
\]
and the diagonal blocks $A$ and $B$ have dimensions $K\times K$ and $(n-K)\times (n-K)$ respectively. In terms of the same type of block matrices the space $\mathfrak{f}_1$ consists of matrices of the type
\[
U_1= \begin{pmatrix}0&C\\D&0\end{pmatrix}.
\]
Since $\sigma$ and $h$ commute the spaces of the real and purely imaginary elements for $\sigma$~-- namely $\su(n)$ and $\ri \su(n)$, are split by $h$ into two subspaces
\begin{gather*}
 \su(n)=(\mathfrak{f}_0\cap \su(n)) \oplus (\mathfrak{f}_1\cap \su(n)), \qquad
 \ri\su(n)=(\mathfrak{f}_0) \oplus (\mathfrak{f}_1).
\end{gather*}
As we shall see $S(x)\in \mathfrak{f}_1$ so the space $\mathfrak{f}_1$ is of particular interest to us. As is easily seen, this space
consists of matrices of the form
\[
R_1= \begin{pmatrix}0&C\\C^{\dag}&0\end{pmatrix},
\]
while $\mathfrak{f}_0\cap \ri\su(n)$ consists of matrices of the form
\[
R_0= \begin{pmatrix}A&0\\0&B\end{pmatrix}, \qquad A^{\dag}=A, \qquad B^{\dag}=B, \qquad \tr A+\tr B=0.
\]
Therefore the real vector space $\mathfrak{f}_1\cap \ri\su(n)$ can be considered as isomorphic to the quotient space $\su(n)\slash ({\rm s}(\u(K)\times \u(n-K)))$.

Below, since all the matrices $S_k$, $S_{k;x}$ belong to $\ri\su(n)$ (that is, they are Hermitian), we shall not write explicitly  that they belong to $\ri\su(n)$ and shall concern ourselves only with whether they belong to $\mathfrak{f}_0$, $\mathfrak{f}_1$, $\mathfrak{h}_S$ and $\mathfrak{h}^{\perp}_S$. This makes the formulas simpler and is possible due to the facts that restricting to $\ri\su(n)$ does not change the form of the recursion operators and the restriction of the Killing form to $\ri\su(n)$ is again nondegenerate.

Now, since $h(S)=-S$ we have $\ad_S\circ h=-h\circ \ad_S$ and the Cartan subalgebra $\mathfrak{h}_S=\ker \ad_S$  also splits into two  subspaces orthogonal to each other
\begin{gather*}
\mathfrak{h}_S={^0\mathfrak{h}}_S\oplus {^1\mathfrak{h}}_S,\qquad
\mathfrak{h}^0_S=\mathfrak{h}_S\cap \mathfrak{f}_0,  \qquad {^1\mathfrak{h}}_S=\mathfrak{h}_S\cap \mathfrak{f}_1.
\end{gather*}
Since orthogonality with respect to the Killing form is preserved by $h$, the space $\mathfrak{h}^{\perp}_S$ is also invariant under $h$ and
\begin{gather*}
\mathfrak{h}_S^{\perp}= {^0\mathfrak{h}^{\perp}_S}\oplus  {^1\mathfrak{h}^{\perp}_S},\qquad
{^0\mathfrak{h}}^{\perp}_S=\mathfrak{h}^{\perp}_S\cap \mathfrak{f}_0, \qquad ^1\mathfrak{h}^{\perp}_S=\mathfrak{h}^{\perp}_S\cap \mathfrak{f}_1.
\end{gather*}

Consider now $S=S_1, S_2,\ldots ,S_{n-1}$, the basis of $\mathfrak{h}_S$ we introduced earlier. From the above it follows that
\[
S_1\in {^1\mathfrak{h}}_S ,\quad S_2\in {^0\mathfrak{h}}_S,\quad S_3\in  {^1\mathfrak{h}}_S,\quad \ldots,
\]
and one can write
\[
S_k\in {^0\mathfrak{h}}_S, \quad k~\mbox{even}, \qquad S_k\in {^1\mathfrak{h}}_S, \quad k~\mbox{odd}.
\]
Consequently,
\[
S_{k;x}\in {^0\mathfrak{h}_S^{\perp}}, \quad k~\mbox{even}, \qquad S_{k;x}\in {^1\mathfrak{h}_S^{\perp}}, \quad k~\mbox{odd}.
\]
In addition, since $\mathfrak{f}_0$ and $\mathfrak{f}_1$ are orthogonal with respect to the Killing form, we have that
\[
\langle S_{2s}, S_{2k-1} \rangle=0.
\]
Finally, one immediately sees that
\begin{gather}\label{eq:exch}
\ad_S({^0\mathfrak{h}_S^{\perp}})={^1\mathfrak{h}_S^{\perp}},\qquad \ad_S({^1\mathfrak{h}_S^{\perp}})={^0\mathfrak{h}_S^{\perp}},\qquad
\ad_S^{-1}({^0\mathfrak{h}_S^{\perp}})={^1\mathfrak{h}_S^{\perp}},\qquad \ad_S^{-1}({^1\mathfrak{h}_S^{\perp}})={^0\mathfrak{h}_S^{\perp}}.
\end{gather}
Let us introduce the following spaces:
\begin{enumerate}\itemsep=0pt
\item $\mathfrak{F}(^0\mathfrak{h}_{S}^{\perp})$ consists of all smooth, rapidly decaying functions $X(x)$ on the line such that $X(x)\in  {^0\mathfrak{h}_{S(x)}^{\perp}}$ (we write simply $X\in {^0\mathfrak{h}_S^{\perp}}$).
\item $\mathfrak{F}(^1\mathfrak{h}_{S}^{\perp})$ consists of all smooth, rapidly decaying functions $X(x)$ on the line such that $X(x)\in  {^1\mathfrak{h}_{S(x)}^{\perp}}$ (we write simply $X\in {^1\mathfrak{h}_S^{\perp}}$).
\end{enumerate}
Naturally,
\begin{gather*}
\mathfrak{F}(\mathfrak{h}_{S}^{\perp})=\mathfrak{F}(^0\mathfrak{h}_{S}^{\perp})\oplus \mathfrak{F}(^1\mathfrak{h}_{S}^{\perp}),
\end{gather*}
where the spaces are orthogonal with respect to the form $\langle\langle\;\cdot\;\rangle\rangle$.
\begin{remark}
Note that all our matrices are also elements from $\ri\su(n)$ so strictly speaking $S_k$ are elements from  $\mathfrak{h}^0_{S}\cap \ri\su(n)$ or $\mathfrak{h}^1_{S}\cap \ri\su(n)$ and $S_{k;x}$ are elements from  $^0\mathfrak{h}_{S}^{\perp}\cap \ri\su(n)$ or $^1\mathfrak{h}_{S}^{\perp}\cap \ri\su(n)$ but we agreed not to write $\ri\su(n)$ as this will make the notation even more complicated.
\end{remark}
After these preliminaries, assuming that all the quantities are as above we have
\begin{proposition}
The recursion operators $\tilde{\Lambda}_{\pm}$ interchange the spaces $\mathfrak{F}(^0\mathfrak{h}_{S})$ and $\mathfrak{F}(^1\mathfrak{h}_{S})$ in the sense that
\[
\tilde{\Lambda}_{\pm}\mathfrak{F}(^0\mathfrak{h}^{\perp}_{S})\subset \mathfrak{F}(^1\mathfrak{h}^{\perp}_{S}), \qquad \tilde{\Lambda}_{\pm}\mathfrak{F}(^1\mathfrak{h}^{\perp}_{S})\subset \mathfrak{F}(^0\mathfrak{h}^{\perp}_{S}).
\]
\end{proposition}
\begin{proof} In order to make the calculation easier, let us introduce the rows of elements:
\begin{gather*}
\mathbf{S}_1=(S_1,S_3,\ldots,S_{2k-1}),\qquad
\mathbf{S}_{1;x}=(S_{1;x},S_{3;x},\ldots,S_{2k-1;x}),
\end{gather*}
where $2k-1$ is the largest odd number less then or equal to $n$ and
\begin{gather*}
\mathbf{S}_0=(S_2,S_4,\ldots,S_{2s}), \qquad
\mathbf{S}_{0;x}=(S_{2;x},S_{4;x},\ldots,S_{2s;x}),
\end{gather*}
where $2s$ is the largest even number less then or equal to $n$.
Let us also introduce the Gram matrices
\begin{gather}\label{eq:Grm0}
^0G= \begin{pmatrix}
\langle S_2,S_2\rangle & \langle S_2,S_4\rangle&\dots& \langle S_2,S_{2s}\rangle\\
\langle S_4,S_2\rangle & \langle S_4,S_4\rangle&\dots& \langle S_4,S_{2s}\rangle\\
\dotfill&\dotfill&\dots&\dotfill\\
\langle  S_{2s},S_2\rangle& \langle S_{2s},S_4\rangle&\dots& \langle S_{2s},S_{2s}\rangle
\end{pmatrix},
\\
\label{eq:Grm1}
^1G= \begin{pmatrix}
\langle S_1,S_1\rangle & \langle S_1,S_3\rangle&\dots& \langle S_1,S_{2k-1}\rangle\\
\langle S_3,S_1\rangle & \langle S_3,S_3\rangle&\dots& \langle S_2,S_{2k-1}\rangle\\
\dotfill&\dotfill&\dots&\dotfill\\
\langle  S_{2k-1},S_1\rangle& \langle S_{2k-1},S_3\rangle&\dots& \langle S_{2k-1},S_{2k-1}\rangle
\end{pmatrix}.
\end{gather}
Since $\langle J_m,J_n\rangle=\langle S_m,S_n\rangle$ these matrices have constant entries. With the help of the matrices~\eqref{eq:Grm0},~\eqref{eq:Grm1} the recursion operators can be written into the form
\[
\tilde{\Lambda}_\pm(\tilde{Z})= \ri  \ad_S^{-1}\pi_S \{ \partial_x\tilde{Z} - \mathbf{S}_{0;x}(^0G)^{-1} \partial_x^{-1}\langle \tilde{Z}_x, \mathbf{S}_0\rangle
-\mathbf{S}_{1;x}(^1G)^{-1} \partial_x^{-1}\langle \tilde{Z}_x, \mathbf{S}_1\rangle\}
\]
or
\[
\tilde{\Lambda}_\pm(\tilde{Z})= \ri  \ad_S^{-1}\pi_S \{ \partial_x\tilde{Z} + \mathbf{S}_{0;x}(^0G)^{-1} \partial_x^{-1}\langle \tilde{Z}, \mathbf{S}_{0;x}\rangle
+\mathbf{S}_{1;x}(^1G)^{-1} \partial_x^{-1}\langle \tilde{Z}, \mathbf{S}_{1;x}\rangle\}.
\]
Assume that $\tilde{Z}\in \mathfrak{F}(^0\mathfrak{h}^{\perp}_{S})$. Then since
$\langle \tilde{Z},\mathbf{S}_0\rangle=\langle \tilde{Z},\mathbf{S}_1\rangle=0$ and
$\langle \tilde{Z},\mathbf{S}_{1;x}\rangle=0$ we get that
\begin{gather}
\tilde{\Lambda}_\pm(\tilde{Z})= \ri  \ad_S^{-1}\pi_S \{ \partial_x\tilde{Z} - \mathbf{S}_{0;x}(^0G)^{-1} \partial_x^{-1}\langle \tilde{Z}_x, \mathbf{S}_0\rangle\}\equiv \tilde{\Lambda}_{\pm}^0(\tilde{Z}),\label{eq:ROpRed0a}\\
\tilde{\Lambda}_\pm(\tilde{Z})= \ri  \ad_S^{-1}\pi_S \{ \partial_x\tilde{Z} +\mathbf{S}_{0;x}(^0G)^{-1} \partial_x^{-1}\langle \tilde{Z}, \mathbf{S}_{0;x}\rangle\}\equiv \tilde{\Lambda}_{\pm}^0(\tilde{Z}).\label{eq:ROpRed0b}
\end{gather}
Recalling \eqref{eq:exch} we get that $\tilde{\Lambda}_{\pm}(\tilde{Z})\in \mathfrak{F}(^1\mathfrak{h}^{\perp}_{S})$. The above also means that trough the expressions~\eqref{eq:ROpRed0a},~\eqref{eq:ROpRed0b} we def\/ine operators $\tilde{\Lambda}_{\pm}^0$ -- the restrictions of $\tilde{\Lambda}_{\pm}$ on $\mathfrak{F}(^0\mathfrak{h}^{\perp}_{S})$. Similarly, if
$\tilde{Z}\in \mathfrak{F}(^1\mathfrak{h}^{\perp}_{S})$ we have
\begin{gather}
\tilde{\Lambda}_\pm(\tilde{Z})= \ri  \ad_S^{-1}\pi_S \{ \partial_x\tilde{Z} - \mathbf{S}_{1;x}(^1G)^{-1} \partial_x^{-1}\langle \tilde{Z}_x, \mathbf{S}_1\rangle\}\equiv
\tilde{\Lambda}_{\pm}^1(\tilde{Z}),\label{eq:ROpRed1a}\\
\tilde{\Lambda}_\pm(\tilde{Z})= \ri  \ad_S^{-1}\pi_S \{ \partial_x\tilde{Z} +\mathbf{S}_{1;x}(^1G)^{-1} \partial_x^{-1}\langle \tilde{Z}, \mathbf{S}_{1;x}\rangle\}\equiv
\tilde{\Lambda}_{\pm}^1(\tilde{Z}),\label{eq:ROpRed1b}
\end{gather}
and we get that $\tilde{\Lambda}_{\pm}(\tilde{Z})\in \mathfrak{F}(^0\mathfrak{h}^{\perp}_{S})$. In the same way as above~\eqref{eq:ROpRed1a},~\eqref{eq:ROpRed1b}
def\/ine operators $\tilde{\Lambda}_{\pm}^1$ -- the restrictions of $\tilde{\Lambda}_{\pm}$ on $\mathfrak{F}(^1\mathfrak{h}^{\perp}_{S})$. In other words we have
\begin{gather*}
\tilde{\Lambda}_{\pm}( \mathfrak{F}(^0\mathfrak{h}^{\perp}_{S}))\in \mathfrak{F}(^1\mathfrak{h}^{\perp}_{S}), \qquad \tilde{\Lambda}_{\pm}( \mathfrak{F}(^1\mathfrak{h}^{\perp}_{S}))\in \mathfrak{F}(^0\mathfrak{h}^{\perp}_{S}).\tag*{\qed}
\end{gather*}\renewcommand{\qed}{}
\end{proof}

In similar situations (when we have  $\mathbb{Z}_2$ reductions) the operators $\tilde{\Lambda}_{\pm}^0$, $\tilde{\Lambda}_{\pm}^1$ are considered as factorizing the recursion operator. In some sense this is true, since if one considers $(\tilde{\Lambda}_{\pm})^2$ acting on $\mathfrak{F}(^0\mathfrak{h}^{\perp}_{S})$  we can write it as $\tilde{\Lambda}_{\pm}^1\tilde{\Lambda}_{\pm}^0$. On the other hand $(\tilde{\Lambda}_{\pm})^2$ acting on $\mathfrak{F}(^1\mathfrak{h}^{\perp}_{S})$  can be written as $\tilde{\Lambda}_{\pm}^0\tilde{\Lambda}_{\pm}^1$. We think that is more accurate to treat the operators $\tilde{\Lambda}_{\pm}^0$, $\tilde{\Lambda}_{\pm}^1$ as restrictions of the operators $\tilde{\Lambda}_{\pm}$. The geometric picture we are going to produce below also supports this viewpoint.

\subsection{Geometric aspects}

The geometric situation in the presence of reductions is also interesting. The point is that the canonical Poisson structure $\bar{Q}_S=\ad_S$ simply trivializes, apparently destroying the
geometric interpretation given in the case of the GZS pole gauge system in general position. Indeed, under the restrictions considered in this section we f\/irst note that the space on which the `point' $S(x)$ takes its values is $\ri\mathcal{O}_J(\SU(n))$ where $J$ is real (and regular). As remarked already this simply makes all the matrices Hermitian and everything remains as it was. However, imposing the second $\mathbb{Z}_2$ reduction (the one def\/ined by the automorphism $h$) means that $S(x)$ belongs to the space of matrices taking values in $\mathfrak{f}_1$ and such that they converge rapidly to some constant values when $x\mapsto \pm\infty$. So for the manifold of potentials $\mathcal{Q}$ we have:
\begin{gather}\label{eq:mfpot}
\mathcal{Q}=\big\{S:  S(x)\in (\ri\mathcal{O}_J(\SU(n)))\cap \mathfrak{f}_1, \ \lim _{x\to \pm\infty}S=S(\pm\infty), \ \mbox{fast enough}  \big\}.
\end{gather}
Then the tangent space $T_S(\mathcal{Q})$ to the manifold \eqref{eq:mfpot} at the point $S$ is the space $\mathfrak{F}(^1\mathfrak{h}^{\perp}_S)$ (for the sake of brevity we again `forget' to mention that the tangent vectors must be also elements of $\ri\su(n)$). But because $\ad_S$ interchanges $\mathfrak{F}(^1\mathfrak{h}^{\perp}_S)$ and $\mathfrak{F}(^0\mathfrak{h}^{\perp}_S)$ and they are orthogonal with respect to $\langle\langle~,~\rangle\rangle$ for  $\tilde{Z}_1,\tilde{Z}_2\in T_S(\mathcal{Q})$ we have that
%\begin{gather*}
$\langle \langle\tilde{Z}_1,[S,\tilde{Z}_2]\rangle\rangle=0$,
%\end{gather*}
so indeed the tensor $\bar{Q}$ becomes trivial. Let us see what happens with the Nijenhuis tensor. Writing everything with the notation we introduced in this section we have
\begin{gather*}
N(X)=\ri\pi_S\partial_x(\ad_S^{-1}X )-\ri\mathbf{S}_{0;x}(^0G)^{-1}\partial_x^{-1}\langle \partial_x(\ad_S^{-1}X), \mathbf{S}_0(x)\rangle\nonumber \\
\hphantom{N(X)}{}
=\ri\pi_S\partial_x(\ad_S^{-1}X )+\ri\mathbf{S}_{0;x}(^0G)^{-1}\partial_x^{-1}\langle \ad_S^{-1}X, \mathbf{S}_{0;x}\rangle =N_0(X) \quad \mbox{for} \ X\in \mathfrak{F}(^0\mathfrak{h}_S^{\perp}), \\
N(X)=\ri\pi_S\partial_x(\ad_S^{-1}X )-\ri\mathbf{S}_{1;x}(^1G)^{-1}\partial_x^{-1}\langle \partial_x(\ad_S^{-1}X), \mathbf{S}_1(x)\rangle  \\
\hphantom{N(X)}{}
=\ri\pi_S\partial_x(\ad_S^{-1}X )+\ri\mathbf{S}_{1;x}(^1G)^{-1}\partial_x^{-1}\langle \ad_S^{-1}X, \mathbf{S}_{1;x}\rangle =N_1(X) \quad \mbox{for} \ X\in \mathfrak{F}(^1\mathfrak{h}_S^{\perp}).
\end{gather*}
Again we can assume that the right-hand sides of the above equations def\/ine the opera\-tors~$N_1$,~$N_2$ and
\begin{gather*}
N_0( \mathfrak{F}(^0\mathfrak{h}^{\perp}_{S}))\in \mathfrak{F}(^1\mathfrak{h}^{\perp}_{S}),\qquad
N_1( \mathfrak{F}(^1\mathfrak{h}^{\perp}_{S}))\in \mathfrak{F}(^0\mathfrak{h}^{\perp}_{S}), \\
N^2\left|_{ \mathfrak{F}(^1\mathfrak{h}^{\perp}_{S})} \right.=N_0N_1,\qquad N^2\left|_{ \mathfrak{F}(^0\mathfrak{h}^{\perp}_{S})} \right.=N_1N_0.
\end{gather*}
An immediate calculation shows that for the conjugates of the operators $N_0$, $N_1$ with respect to $\langle\langle~,~\rangle\rangle$ we have
\[
N_0^*=\tilde{\Lambda}^0_{\pm}, \qquad N_1^*=\tilde{\Lambda}^1_{\pm}.
\]
\begin{remark}
One should bear in mind that $\partial_x^{-1}$  that enters in the expressions for $N_0$, $N_1$ is treated either as $\displaystyle \int_{-\infty}^x  \cdot\,\rd y$ or as $\displaystyle \int_{+\infty}^x \cdot\, \rd y$.
\end{remark}
One can easily see that $N(T_S(\mathcal{Q}))$ does not belong to $T_S(\mathcal{Q})$, so the restriction of $N$ on $\mathcal{Q}$ cannot be a Nijenhuis tensor on $\mathcal{Q}$. However, $N^2(\mathfrak{F}(^1\mathfrak{h}^{\perp}_{S}))\subset\mathfrak{F}(^1\mathfrak{h}^{\perp}_{S})$, that is
\[
N^2(T_S(\mathcal{Q}))\subset T_S(\mathcal{Q}),
\]
so $N^2$ becomes a natural candidate. Indeed, let us recall the following facts from the theory of P-N manifolds, see
\cite{GerViYa2008, Magr80,Magr78,MagMor*84}:
\begin{theorem}
If $\mathcal{M}$ is a P-N manifold endowed with Poisson structure $P$ and Nijenhuis tensor~$N$, then for $k=1,2,\ldots$ each pair $(N^kP=P(N^*)^k, N^s)$ also endows $\mathcal{M}$ with a P-N structure.
\end{theorem}

\begin{theorem}
Let $\mathcal{M}$ be a P-N manifold endowed with Poisson tensor $P$ and Nijenhuis tensor~$N$. Let $\bar{\mathcal{M}}\subset \mathcal{M}$ be a submanifold of $\mathcal{M}$ and suppose that we have:
\begin{enumerate}\itemsep=0pt
\item [i)] $P$ allows a restriction $\bar{P}$ on $\bar{\mathcal{M}}$ such that if $\rj:\bar{\mathcal{M}}\mapsto \mathcal{M}$ is the inclusion map then $\bar{P}$ is $\rj$-related with $P$, that is $P_m=\rd\rj_m\circ\bar{P}_m\circ (\rd\rj_m)^*$,   $m\in \bar{\mathcal{M}}$.
\item[ii)] The tangent spaces of $\bar{\mathcal{M}}$, considered as subspaces of the tangent spaces of $\mathcal{M}$ are invariant under $N$, so that $N$ allows a natural restriction $\bar{N}$ to $\bar{\mathcal{M}}$, that is $\bar{N}$ is $\rj$-related with $N$.
\end{enumerate}
Then $(\bar{P},\bar{N})$ endow $\bar{\mathcal{M}}$ with a P-N structure.
\end{theorem}
We call the above theorem the \emph{second restriction theorem}.

 In view of what we have already, we need only f\/ind the restriction $\bar{P}'$ of $\bar{P}$ on $\mathcal{Q}$ and then~$\bar{P}'$ and the restriction of~$N^2$ will endow $\mathcal{Q}$ with a P-N structure.
Thus we have the following candidates for restriction -- the Poisson tensor $\bar{P}=N\circ \ad_S=\ad_S\circ N^*$ and the Nijenhuis tensor~$N^2$ (or~$N^{-2}$).
Let us take $\bar{P}$ and try to restrict it.

We want to apply the f\/irst restriction theorem. $\mathcal{X}^*(\bar{P})_S$ consists of smooth functions $\beta$, going rapidly to zero when $|x|\to \infty$ such that $\beta(x)\in \mathfrak{h}_S^{\perp}(x)$ and  $\bar{P}(\beta)\in T_S(\mathcal{Q})$. The last means that
\[
\ri\pi_S\partial_x\beta-\ri\mathbf{S}_xG^{-1}\partial_x^{-1}\langle \partial_x\beta(x), \mathbf{S}(x)\rangle\in \mathfrak{f}_S^{1}(x),
\]
that is, for arbitrary smooth function $X(x)$ such that $X\in \mathfrak{f}_S^{0}$ and going rapidly to zero when $|x|\to \infty$ we have
\begin{gather}\label{eq:con1}
\langle \ri\pi_S\partial_x\beta-\ri \mathbf{S}_xG^{-1}\partial_x^{-1}\langle \partial_x\beta(x), X(x)\rangle=0.
\end{gather}
The space $T^{\perp}(\mathcal{Q})_S$ consists of smooth $\beta(x)$ such that $\beta\in \mathfrak{h}_S^{\perp}$, going rapidly to zero when $|x|\to \infty$ and satisfying
\[
\langle\langle \ri\pi_S\partial_x\beta-\ri\mathbf{S}_x G^{-1}\partial_x^{-1}\langle \partial_x\beta(x), \mathbf{S}(x)\rangle, Y(x)\rangle\rangle=0
\]
for each smooth function $Y(x)$, $Y\in \mathfrak{f}^1_S(x)$ going rapidly to zero when $|x|\to \infty$. Arguments similar those used to prove Haar's lemma in the variational calculus show that for each $x$
\begin{gather}\label{eq:con2}
\langle \ri\pi_S\partial_x\beta-\ri \mathbf{S}_xG^{-1}\partial_x^{-1}\langle \partial_x\beta(x), \mathbf{S}(x)\rangle, Y(x)\rangle=0,
\end{gather}
where $Y(x)$ is as above. Then, if $\beta\in  \mathcal{X}^*(\bar{P})_S\cap T^{\perp}(\mathcal{Q})_S$, we shall have simultaneously~\eqref{eq:con1} and~\eqref{eq:con2} so  $\beta\in \ker \bar{P}_{S}$. Thus the f\/irst condition of the f\/irst restriction theorem is fulf\/illed. In order to see that the second  condition also holds, we introduce
\begin{lemma}
The operator $\bar{P}$ has the properties
\[
\bar{P}_{S}(\mathfrak{F}(\mathfrak{f}^0_S))\subset \mathfrak{F}(\mathfrak{f}^0_S),\qquad \bar{P}_{S}(\mathfrak{F}(\mathfrak{f}^1_S))\subset \mathfrak{F}(\mathfrak{f}^1_S).
\]
\end{lemma}
The proof of the lemma is obtained easily since the spaces $\mathfrak{f}^0_S$ and $\mathfrak{f}^1_S$ are invariant with respect to  $\pi_S$.

Using the lemma, suppose $\beta\in \mathfrak{h}_S^{\perp}$ is a smooth function  going rapidly to zero when $|x|\to \infty$. Clearly, we can write it uniquely into the form
\[
\beta=\beta_0+\beta_1,\qquad\beta_0\in \mathfrak{F}(\mathfrak{f}^0_S),\qquad \beta_1\in \mathfrak{F}(\mathfrak{f}^1_S).
\]
Then $\bar{P}_{S}(\beta_0)\in \mathfrak{F}(\mathfrak{f}^0_S)$, $\bar{P}_{S}(\beta_1)\in \mathfrak{F}(\mathfrak{f}^1_S)$ and we see that $\bar{P}_{S}\beta_0\in \mathcal{X}^*(\bar{P})_S$, $\bar{P}_{S}\beta_1\in T^{\perp}(\mathcal{Q})_S$. So the second requirement of the f\/irst restriction theorem is also satisf\/ied and $\bar{P}$ allows restriction. If~$\gamma$ is a $1$-form on $\mathcal{Q}$, that is $\gamma\in \mathfrak{F}(\mathfrak{f}^1_S)$, the restriction $\bar{P}'$ is given by
\begin{gather}\label{eq:Poisson}
\bar{P}'(\gamma)=\ri\pi_S\partial_x\gamma-\ri \mathbf{S}_{1;x}(^1G)^{-1}\partial_x^{-1}\langle \partial_x\gamma(x), \mathbf{S}_1(x)\rangle.
\end{gather}
 We summarize these facts into the following
\begin{theorem}
The manifold of potentials $\mathcal{Q}$ is endowed with a P-N structure, defined by the Poisson tensor $\bar{P}'$ \eqref{eq:Poisson} and the restriction of the Nijenhuis tensor $N^2$. Explicitly, the restriction of $N^2$ is given by $N_0N_1$ where:
\begin{gather*}
N_0(X)=\ri\pi_S\partial_x(\ad_S^{-1}X) -\ri\mathbf{S}_{1;x}(^1G)^{-1}\partial_x^{-1}\langle \partial_x(\ad_S^{-1}X), \mathbf{S}_1(x)\rangle, \\
\hphantom{N_0(X)=}{} \
X\in \mathfrak{F}(\mathfrak{f}_S^{0}),\qquad  N_0(X)\in \mathfrak{F}(\mathfrak{f}_S^{1}),  \\
N_1(X)=\ri\pi_S\partial_x(\ad_S^{-1}X )-\ri\mathbf{S}_x\partial_x^{-1}\langle \partial_x(\ad_S^{-1}X), \mathbf{S}_0(x)\rangle , \\
\hphantom{N_1(X)=}{} \
X\in \mathfrak{F}(\mathfrak{f}_S^{1}),\qquad  N_0(X)\in \mathfrak{F}(\mathfrak{f}_S^{0}).
\end{gather*}
 \end{theorem}
Since $(N_0)^*=\Lambda_{\pm}^0$, $(N_1)^*=\Lambda_{\pm}^1$ the above theorem gives a geometric interpretation of the operators $\Lambda_{\pm}^{0,1}$.

Finally, let us calculate the Poisson bracket for the Poisson structure $\bar{P}'$. If~$H_1$,~$H_2$ are two allowed functionals and we introduce the row $\langle \mathbf{S}_{1;x},\frac{\delta H_2}{\delta S}\rangle^t(x)$ with elements  $\langle S_{r;x},\frac{\delta H_2}{\delta S}\rangle(x)$, $r\leq n$, $r=1,3,5,\dots$, we shall obtain
\begin{gather*}
\{H_1,H_2\}=
\ri\left\langle\left\langle \partial_x\left(\frac{\delta H_1}{\delta S}\right),\frac{\delta H_2}{\delta S}\right\rangle\right\rangle \\
\hphantom{\{H_1,H_2\}=}{}
+\frac{\ri}{2}\int_{-\infty}^{+\infty}\left[\left\langle \mathbf{S}_{1;x},\frac{\delta H_2}{\delta S}\right\rangle^t(x) (^1G)^{-1} \left(\int_{-\infty}^x + \int_{+\infty}^x\right) \left\langle \mathbf{S}_{1;y},\frac{\delta H_1}{\delta S}\right\rangle(y) \rd y\right]\rd x.
\end{gather*}

\subsection{The hierarchies of integrable equations}

The geometric theory is incomplete without giving (at least partially) the hierarchies of equations (vector f\/ields) that correspond to the P-N structure described in the above theorem, that is the f\/ields that are fundamental  for the P-N structure.  In order to discuss this issue let us start with the fundamental f\/ields for the P-N structure in the general case (without reductions). As it is known, see~\cite{GerViYa2008},  the f\/ields of the type $X_H(S)=[H, S]$, $H\in \mathfrak{h}$ are fundamental for the P-N structure. This is because the original tensors~$P$,~$Q$ (see~\eqref{eq:P}) are covariant with respect to the one-parametric group of transformations
\[
\varphi^H_t(S)=\Ad(\exp{tH})S, \qquad H\in \mathfrak{h},  \qquad t\in \mathbb{R},
\]
and the submanifold $\mathcal{N}$ is invariant with respect  to $\varphi^H_t$. The one-parametric group $\varphi^H_t$ corresponds to the vector f\/ields $X_H$ which are tangent to $\mathcal{N}$.  According to the theory of the P-N manifolds, if $X$ is a fundamental f\/ield then all the f\/ields $N^pX$, $p=1,2,\ldots$ are also fundamental and commute, that is they have zero Lie brackets. Thus to each $H\in \mathfrak{h}$ corresponds a hierarchy of fundamental f\/ields $N^k[H, S]$ and we obtain $n-1$ independent families of fundamental f\/ields. For dif\/ferent $p$ and dif\/ferent  $H\in \mathfrak{h}$ the f\/ields from these families also commute.  As a matter of fact this is commonly referred to as `the geometry' behind the properties of the hierarchies~\eqref{eq:EEZSSspole}.  We can cast these hierarchies in a dif\/ferent form.  Using the expression~\eqref{eq:NGP1} and taking into account that $\ad_S^{-1}\ad_SH=\pi_S(H)$ after some simple transformations in which we use the properties of the Killing form and  Proposition \ref{prop:ort} we get
\begin{gather}
N(X_H)(S)=\ri\pi_S\partial_x(\pi_S(H))-\ri\mathbf{S}_xG^{-1}\partial_x^{-1}\langle \partial_x(\pi_S(X)), \mathbf{S}(x)\rangle\nonumber\\
\hphantom{N(X_H)(S)}{}
=-\ri\mathbf{S}_x G^{-1}\langle H,\mathbf{S}(\pm\infty)\rangle=-\mathbf{S}_x G^{-1}\langle H,\ri\mathbf{S}(\pm\infty)\rangle. \label{eq:hiergen}
\end{gather}
The last expression shows that $N(X_H)(S)$ is a linear combination of the f\/ields $S_{1;x}=S_x$, $S_{2;x}=(S^2)_x,\ldots, S_{n-1;x}=(S^{n-1})_x$. Since the evolution equations corresponding to the f\/ields~$X_H$ are linear one can say that the hierarchies of the fundamental f\/ields that correspond to nonlinear equations are generated by  the above vector f\/ields.  Now, let us assume that one has reductions as in the above and let us f\/irst consider what happens when we restrict to~$\ri \su(n)$.  In that case we must take $H\in \su(n)$, that is, $\ri H$ must be a diagonal, traceless real matrix. Now we can see that the
coef\/f\/icients in front of the f\/ields $S_{k;x}$ are real since $\su(n)$ is a compact real form of~$\asl(n)$ and on it the Killing form is real so both $G$ and $\langle H,\ri\mathbf{S}(\pm\infty)\rangle$ are real.

When we have the additional restriction def\/ined by the automorphism $h$ the situation is more complicated. Since $h(S)=-S$ and $h(H)=H$ for any $H\in \mathfrak{h}$ we have
$h(X_H)=-X_H$ so   $X_H\in  \mathfrak{F}(^1\mathfrak{h}_{S}^{\perp})$. Thus $N(X_H)\in  \mathfrak{F}(^0\mathfrak{h}_{S}^{\perp})$  and more generally for $p=0,1,2,\ldots$
\begin{gather} \label{eq:series}
N^{2p}X_H= (N_0N_1)^pX_H    \in  \mathfrak{F}(^1\mathfrak{h}_{S}^{\perp}),\qquad
N^{2p+1}X_H= (N_1N_0)^pN_1X_H    \in  \mathfrak{F}(^0\mathfrak{h}_{S}^{\perp}).
\end{gather}
\begin{remark}
The fact that  $N(X_H)\in  \mathfrak{F}(^0\mathfrak{h}_{S}^{\perp})$  can also be seen easily from~\eqref{eq:hiergen}. Indeed,  because $H$ is an element of the Cartan subalgebra it is orthogonal to all elements of $\mathfrak{f}_1$  and in~\eqref{eq:hiergen} we have a linear combination of the elements $S_m$ with odd index $m$, that is $ -\mathbf{S}_x G^{-1}\langle H,\ri\mathbf{S}(\pm\infty)\rangle=-\mathbf{S}_{1;x} {^1G^{-1}}\langle H,\ri\mathbf{S}_{1}(\pm\infty)\rangle$.
\end{remark}
Thus when we have a reduction def\/ined by the automorphism $h$  the f\/irst of the series in \eqref{eq:series} consists of vector f\/ields tangent to the manifold $\mathcal{Q}$ while the vector f\/ields from the second series are not tangent to~$\mathcal{Q}$.

Naturally, the vector f\/ields from the f\/irst series are the candidates for being fundamental f\/ields of the P-N structure on~$\mathcal{Q}$ and, discarding the f\/irst f\/ields in the hierarchies, we see that we have f\/ields of the type
\[
N^{2p}S_{2j-1;x}= (N_0N_1)^pS_{2j-1;x}.
\]
These are not, however, all the fundamental f\/ields that we can produce.
The f\/ields of the type $NS_{2j;x}$ are fundamental for $N$ and hence are fundamental for $N^2$ also. Besides, since  $S_{2j;x} \in  \mathfrak{F}(^0\mathfrak{h}_{S}^{\perp})$ we have $NS_{2j;x} \in  \mathfrak{F}(^1\mathfrak{h}_{S}^{\perp})$. As easily checked,  the f\/ields $NS_{2j;x}$ are tangent to $\mathcal{Q}$.

Thus, f\/inally, the hierarchies of the  the fundamental f\/ields of the P-N structure when we restrict to  $\mathcal{Q}$ are generated by the f\/ields
\begin{gather*}
N^{2p}S_{2j-1;x}=(N_0N_1)^pS_{2j-1;x}, \qquad N^{2p+1}S_{2l;x}=(N_0N_1)^pN_0S_{2l;x},\\
1\leq j\leq k, \qquad 1\leq l\leq s, \qquad p=0,1,2,\ldots,
\end{gather*}
where $k$ and $s$ are such that $2k-1$ is the largest odd number less than $n$ and $2s$ is the largest even number less than $n$. These numbers def\/ine the sizes of the Gram matrices $^0G$ and $^1G$.  For example, in the case of the algebra $\asl(3,\mathbb{C})$, we have the f\/ields
\begin{gather}
N^{2p}S_{x}=(N_0N_1)^p S_{x}, \qquad N^{2p+1}(S^2)_{x}=(N_0N_1)^pN_0(S^2)_{x},\qquad p=0,1,2\ldots.\label{eq:ff}
\end{gather}

These fundamental f\/ields give rise to hierarchies of integrable equations of the type $S_t=F(S)$ where $F(S)$ is a f\/inite linear combination of the fundamental f\/ields \eqref{eq:ff}, see \cite{GerGrahMikhVal*2012}.

\section{Conclusions}
We have been able to show that the geometric interpretation known for the the recursion opera\-tors related to the generalized Zakharov--Shabat system on the algebra $\asl(n,\mathbb{C})$ in pole gauge holds also in the presence of $\mathbb{Z}_2\times \mathbb{Z}_2$ reductions of certain classes and we have explicitly calculated the recursion operators. It is an interesting question whether analogous results could be obtained for Zakharov--Shabat type systems on the other classical Lie algebras.

\subsection*{Acknowledgements}

The authors are grateful to A.V.~Mikhailov and V.S.~Gerdjikov for drawing their attention to the theory of the recursion operators in the presence of reductions and the problems related to it. We would like also to thank the referees who read carefully the manuscript and make number of remarks that helped to improve considerably the manuscript.

%\cite{G86,GerGrahMikhVal*2012,GKS,Magr78,Magr80,Yan2012b}

\pdfbookmark[1]{References}{ref}
\LastPageEnding

\end{document}